# Exploring Peer Review in the Computer Science Classroom

Scott Turner

Manuel A. Pérez-Quiñones

(ABSTRACT)

In computer science, students could benefit from more opportunities to learn important, high-level concepts, to perform professional activities, and to improve their learning skills.  Peer review is one method to encourage this but it is not commonly used in the discipline.  Peer review provides students with both the opportunity to evaluate other people's work, to receive additional feedback on their own projects, and, in doing so, allows for rich learning experience.  However, when reviewing a programming project, it is not immediately obvious how to create an activity that will improve the students' understanding of the concepts, require higher level thinking, and be engaging for a particular class.  The current literature does not address this issue well.  There are few comparisons between review implementations and few reasons for design decisions.

The goal of this work was to explore the questions of what, who, and when to review.  Specifically, this involved manipulating the *materials* reviewed, the *type* of review, the class *level* of the students, and the *position* of the review within the project to determine how these dimensions changed the students' learning of concepts, high-level thinking, and engagement.  A study involving multiple reviews across many different classes was conducted to examine the effects of the different implementations.

Unfortunately, due to a large amount of incomplete data, only the *type* of review could be reasonably analyzed.  From the analysis results, there were two interesting findings.  First, between reviews of different *types*, there were changes in how well students addressed the concept they were reviewing (i.e. they talked about Decomposition when asked about the project's Decomposition) and in the length of the comments.  This is an indication that the review's *type* may affect student engagement and the how well they understand or are learning the concepts.  Second, there were differences in *how* they reviewed the concepts of Abstraction, Decomposition, and Encapsulation.  This suggests that the concepts are understood or are being learned in different ways.  Both of these results have an impact on how peer review is used in computer science but need further investigation.  This work concludes with a number of questions to be answered with additional research.

# 1 Introduction

Like all disciplines, when teaching computer science, some of the educational goals are to expose students to concepts in the discipline, to prepare them to perform professional activities and to teach them how to learn.  There are numerous ways to reach these objectives such as having students read textbooks or research papers, create class designs, write programs, or use the tools of the trade.  Another way is to have the learners read code and designs by reviewing the work of their peers.

Peer review, while it has many known benefits (Zeller 2000; Papalaskari 2003; Wolfe 2004; Hamer, Ma et al. 2005; Trytten 2005), and is used extensively in other fields (Falchikov and Goldfinch 2000; Topping, Smith et al. 2000; Liu and Hansen 2002; Dossin 2003; Carlson and Berry 2005) and in industry (Anderson and Shneiderman 1977; Anewalt 2005; Hundhausen, Agrawal et al. 2009), is not as widely used in the computer science curriculum.  This may be due, in part, to a lack of information about *what*, *who*, and *when* to review in order to achieve specific goals in computer science.  That is not to say that the literature is silent on these issues.  The studies make these choices but there are few reasons given for the decisions and even fewer comparisons performed to show relative value of those options.  What is needed is a clearer understanding of the requirements that the discipline imposes on the peer review process so that it may be effectively used.

Peer review is, potentially, a very worthwhile learning endeavor.  Peer review's two main activities, the giving and receiving of feedback, involve the active use of the higher levels of Bloom's taxonomy, synthesis, analysis, and evaluation (Bloom 1984) and can provide students with a rich opportunity for learning.  When students are critiquing another person's work, they are, of course, evaluating the quality of the work, but they are also analyzing the choices and patterns made and they are synthesizing them into their own body of knowledge.  In addition, they get the chance to see some of the concepts they have been taught put to use in an authentic situation.  So, not only do they have the opportunity to learn from the assessment of the work, but from the new ideas, choices, and techniques (good or bad) that they see (Papalaskari 2003; Trytten 2005).

Besides the benefits to learning, this is also a necessary skill for students to have as it prepares them for professional activity.  Being able to read, understand, and evaluate code is needed throughout the stages of a program's life cycle (Deimel 1985).  For instance, when writing or maintaining programs, one must be able to look at what has been done in order to build on it, modify it, or correct it.  That is true if for small, individual school work and for large, group-developed, professional products.

Looking at the other side, receiving feedback from others is an activity of synthesis, analysis, and evaluation.  Students go over the comments of the reviewers and look for patterns and try to add them to their own understanding of their project.  There is also some evaluation going on as the students compare the feedback against the guidelines and the project in an effort to determine if they are valid and worthwhile comments.

As said, this is *potentially* a very beneficial activity.  Computer science provides a different environment in which to review and that must be considered as these activities are created and run.  One must consider what goals should be met by the assignment.   One implementation of the review may be better for improving the quality of the projects than for increasing student motivation and self-efficacy and vice versa.  Understanding what should be achieved and the best ways to reach those goals is important to improving learning.

*What* is being reviewed could affect the outcomes.  The *materials* (design diagrams, code, etc.) lend themselves to different approaches for reviewing.  A term paper is not read the same way that code is.  Add to that the reality that students may not be familiar with reading code and it becomes plain that the kind of object being reviewed may alter the review and its effects.

*Who* is being reviewed is a similar issue.  When considering the *type* of review, in a peer review, the students are reviewing their classmates' materials and, in a training review, the students are reviewing the instructor's.  Reviewing peers may provide feedback to other students but it may not provide everyone with the same types of situations to address.  Having the instructor provide the material to review would make the process more uniform but would lose the interaction between the students.  Those effects may be more or less beneficial depending on the situation.

*When* to review occurs has two parts to address.  First, there is the level of the class.  What freshmen need to learn and experience is different from what seniors or graduate students need.  Certain implementations of the review process may have more or less effect depending on the class.  Second, there is the question of when in the course of a project's development the review should happen.  Early, formative reviews would allow students to make changes to their plans.  Late, summative reviews would allow students to see a completed, alternate approach to a project.

These questions form just a part of peer review's enormous design space.  To help frame our understanding of this space, Topping provides us with a topology of 17 dimensions in which to work (See 2.1.2 Peer Review Theories and Frameworks) (Topping 1998).  These include the *Objective* for the review, its *Directionality*, its formative or summative *Focus*, the *Year* of the class, the *Product/Output* under review, and other such concerns.  This framework allows us to describe our review assignments while exploring their benefits.

Using these questions and this framework, we want to explore the issues so that we can provide guidance in the design of peer review exercises.  This is a complicated process and one that is not fully understood in computer science.  The more that is known about the relative effects of these design choices, the better we can plan for them.

## 1.1 Research Goals

For this work, we are exploring the peer review process in computer science.  We are looking at its use in computer science programming classes as a way to promote:

- The understanding of high-level object-oriented programming concepts (Henderson-Sellers 1992; Armstrong 2006)

- Higher-level thinking (Bloom 1984)

- Active, engaged learning (Huba and Freed 1999)

To reach these goals, we are comparing implementations of the review process within and between classes.  We are manipulating the

- *Materials* under review (design diagrams versus code)
- *Type* of review (training with instructor material versus peer reviews)
- *Level* on which the review is completed (CS1 versus CS2 versus …)
- *Position* of the review within the  project (early (formative) versus late (summative))

From the results of the exploration, we can better describe how to approach the creation of a review for specific goals in a specific setting.

The next section examines relevant research in the area and explores some of the theories related to peer review.   Following that is a discussion of our methods and results.  Finally, there is a discussion of the findings and future work.

# 2 Literature Review

To understand the peer review task in the context of computer science, it is, of course, necessary to understand how the peer review process functions in other fields. It is also necessary to understand the theories behind it and to adopt a framework that will allow us to describe the aspects of peer review that we are interested in. Not only do we need to be able to create and execute peer reviews, but we need to do so in way that focuses on the student and does not distract from the learning goals.

This chapter first covers some of the aspects of peer review. This includes some of the benefits of the process and identifies some common concerns, such as anonymity, quality of reviews and assignment of reviewers, which are found throughout the literature. After that are discussions of rubrics, of context and the role it plays and of the use of technology in the classroom.

## 2.1 Peer Review

In computer science, there is large amount of literature about the benefits of peer review and about students' reactions to the process. There have also been a number of studies involving the creation of peer review systems and the viability of computer mediated reviewing. From the literature, we have seen, among others, studies that have evaluated:

- How closely student reviews matched other students or instructor reviews (Falchikov and Goldfinch 2000; Gehringer 2001; Liu, Lin et al. 2002; Hamer, Ma et al. 2005; Sadler and Good 2006; Tseng and Tsai 2007; van Hattum-Janssen and Lourenco 2008)

- Basic student motivation and attitudes towards peer review (Liu and Yuan 2003; Li 2006; van Hattum-Janssen and Lourenco 2008; Xiao and Lucking 2008)

- Systems for performing reviews (Mathews and Jacobs 1996; Preston and Shackelford 1999; Davies 2000; Silva and Moriera 2003; Trivedi, Kar et al. 2003; Sitthiworachart and Joy 2004; Powell, Turner et al. 2006; Denning, Kelly et al. 2007; Wang, Yijun et al. 2008; Xiao and Lucking 2008)

- How well students identified problems in code (Chinn 2005)

- The usefulness of expert vs. novice reviews (Cho, Chung et al. 2008)

- Improvements in student projects (Tseng and Tsai 2007; Hundhausen, Agrawal et al. 2009)

The approaches and the corresponding the systems found in much of the literature make design decisions (supporting or not supporting anonymity, use of rubrics or in-document annotations, etc.) based on some notion of how the reviews should be structured (Joy and Luck 1998; Mason and Woit 1999; Preston and Shackelford 1999; Zeller 2000; Gehringer 2001; Liu and Yuan 2003; Silva and Moriera 2003; Trivedi, Kar et al. 2003; Wolfe 2004). While these may be legitimate ways of approaching the problem, there are few reasons for the design choices and few explanations of why this implementation is best for this situation. Is reviewing code a good way to promote learning in computer science? Would using designs be a reasonable alternative?  Do the course and its topics make a difference in the review's effectiveness? Do students learn better if they are reviewing work from other students in the class or something provided by the instructor?  These and many more questions have yet to be answered.

In this work, we want to focus on peer review as a way to improve students' understanding of high-level object-oriented programming concepts, to increase higher-level thinking, and to raise engagement with the material.  To do that, we want to explore different implementations of peer review to compare the effects of using different materials (designs or code), reviewing different people (peers or the instructor), evaluating in different courses (CS1 or CS2 or …), and changing the timing of the feedback (formative or summative).

### 2.1.1 Motivation

One reason for using peer review in the classroom is that it is an active learning exercise (McConnell 2001) that offers a different type of learning activity.  This has the benefit of breaking the monotony of a class in a typical lecture format.  It also varies style in which the class is taught (Felder and Silverman 1988; Felder, Felder et al. 2002).  Since a class will have students with a variety of learning styles, changing the method of teaching will spread the benefits to more of the class.  It also has the effect of giving people a chance to improve how they learn in a style that they do not prefer.  This may have the effect of making them more well-rounded learners.

Peer reviews can be developed to address a number of learning styles.  They can be used to promote the active and/or reflective processing of information.  Discussion can be created around the reviews or, at the very least, the person being reviewed receives a one-sided conversation to consider.  Those that want to be more reflective have the opportunity to look at and compare themselves to others.  Depending on what is being reviewed, say a diagram or an essay, visual or auditory styles may be helped.  The instructions given of the review itself could make the process more inductive or deductive.  Students might be told to look for a number of specific qualities from which they could infer what makes a good project or paper.  Alternatively, they might be told to give marks based on style from which they could develop their own criteria.  There are many possibilities to be explored.

### 2.1.2 Peer Review Theories and Frameworks

There are several educational theories that can be used to explore the idea of peer review.  One way to look at the process is from a cognitive viewpoint.  Bloom's taxonomy (Bloom 1984) can be used for this.  Peer review supports higher-level learning skills such as synthesis, analysis, and evaluation.  Students are put into a situation where they must judge someone else's work.  This provides them with the opportunity to compare it with their own work and to think about how some of the choices made are applicable in other situations.  There may also be some learning on lower-levels, such as knowledge and understanding, if the students see new uses of standard classes, methods, or operators that they have not seen before.

Piaget's work (1950) provides a constructivist approach to peer review.  (Wankat and Oreovicz (1992) provides a brief overview of Piaget's theories.)  Peer review is an active process that engages the students in learning and gives them some ownership in their instruction.  Having access to other peoples' work can provide new perspectives that challenge the student's.  In addition, the feedback provided to the student may also cause enough cognitive dissonance to encourage the student to modify his or her concepts of the task.  This helps the student identify incorrect facts and other fallacies so that new learning can commence.

Related to constructivism are the ideas behind Problem-based Learning (Ellis, Carswell et al. 1998).  Reading, understanding, and evaluating other peoples' work are common in professional settings (Deimel 1985).  Peer review could be used to develop these skill in an authentic (or, at least, a semi-authentic) manner.

Social constructivism is another way in which peer review can be approached.   Vygotsky's Zone of Proximal Development (Doolittle 1997) suggests that the interaction between reviewer and reviewee allows for the creation of knowledge.  The differences in knowledge, points of view, and experiences should provide an opportunity for dialog and for negotiation of meaning.  By working together in this way, students will be able to accomplish and learn more than they could alone.

The educational and psychological theories that apply to peer review are not limited to those mentioned above.  Perry's work on mental development could also be applied (Perry 1970).  (Wankat and Oreovicz (1992) also provides a brief overview of Perry's theories.)  Other theories, such as Process Writing Theory and Collaborative Learning Theory, support the same ideas (Liu and Hansen 2002).  The peer review process is complex enough that many theories may be appropriate and, yet, they may not completely capture it.

These different approaches to describing peer review appear throughout the literature.  While many authors referred to social constructivist ideas, specifically Vygotsky, there were a number of other theories mixed in as well.  For example, Topping's work (Topping 1998; Topping, Smith et al. 2000; Topping 2005) draws from both Piaget and Vygotsky.  Falchikov and Goldfinch's (Falchikov and Goldfinch 2000) meta-analysis of peer review studies used a similar approach.  King (1998), whose work on peer tutoring shares many attributes with peer review, also uses Vygotsky as a foundation.  Chinn (2005) grounds his ideas using the work of, among others, Bloom, Vygotsky, and Perry.  Both Sitthiworachart and Joy (2004) and Trytten (2005) also use ideas from Bloom and, while not clearly expressed, there are social constructivist aspects as well.

That being said, it would seem that the kind of peer review being done greatly influences what theory would best describe the process.  That is, a set of peer reviews done over an extended length of time, within the same groups, where there is considerable discussion, and where there is no anonymity may be best viewed from a social constructivist standpoint.  There is much more time for interaction between the peers and for the building of knowledge.  On the other hand, a one-time review where the discussion is one-way and anonymous may lend itself more to a cognitive approach.  The social aspect of the process is greatly deemphasized and more attention can be paid to the changes in the students' thought processes.

Since peer reviews can be carried out in so many different ways, a way to describe and classify the process is needed. Thankfully, Topping (1998) provides a broad topology of the peer reviews process along 17 dimensions which include directionality (one-way, reciprocal, mutual), privacy, and value of the review (grade or other). (See Table 1 for more details.) This provides a solid foundation to explore the peer review activity within.

**Table 1: Topping's Peer Review Topology (Topping 1998)**

| Dimension | Description |
| --- | --- |
| **Curriculum Area** | The subject area in which the review takes place. This could be any subject. |
| **Objectives** | The reason for performing the review. This could be to provide more feedback, to reduce the teacher's grading load, to allow the students to see other possible solutions, etc. Students and teachers may or may not have similar objectives. |
| **Focus** | The aim of the review. Is it providing formative or summative feedback? Is it quantitative or qualitative? |
| **Product/Output** | The work being reviewed. This could be almost anything including written papers, programs, diagrams, presentations, performances, forum discussions and graded work. |
| **Relation to Staff Assessment** | The overlap between peer and instructor assessment. The review could be in addition to Instructor and TA feedback or it could be the only feedback provided. |
| **Official Weight** | The amount this review effects the reviewee's grade. |
| **Directionality** | The logic of who is reviewed by whom. This could be one-way, where a student gives a review but is not reviewed, reciprocal, where pairs could review each other, or mutual, where everyone gives and receives a review from someone in the group. |
| **Privacy** | The amount of anonymity in the review. Do one or both the parties know the identity of other? Does the general public? Are the reviews made available to the teacher, the class, or everyone? |
| **Contact** | The interaction between reviewer and reviewee. This could be face-to-face, mediated online, or done through other communication devices. There may be discussion between the parties or there may be no interaction except through the work being reviewed and the review. |
| **Year** | The relative grade level of the students. Are students matched to those in the same grade, across grades, or both? |
| **Ability** | The relative skills of the students. Are students matched to those with a similar level of ability, across levels, or both? |
| **Constellation Assessors** | The structure of the reviewers. Is the review done by a single student or by multiple students? |
| **Constellation Assessed** | The structure of the reviewees. Is the review done for a single student or for multiple students? |
| **Place** | The location of the review. This could be in and/or out of class. |
| **Time** | The time spent on the review. This could be in and/or out of class. |

| Requirement | The official mandate for participating in the review. Is reviewing required? Is submitting work to be reviewed required? |
|---|---|
| **Reward** | The benefits associated with completing a review. This could include class credit, recognition, etc. |

As can be seen from the topology, there is a multitude of ways to implement a review. Reviewers and reviewees can be given differing amounts of motivation and benefits by adjusting some of the dimensions. This implies that one can tailor the activity for specific goals. It also highlights some of the aspects that must be considered during the planning and creation of peer reviews. If reviews are being done across ability levels, is there enough of a reward built into the system that the student with the higher ability feels that he or she is getting some benefit from doing the review? If the review is formative and qualitative, should it count towards the final grade? The framework provides some structure when considering these types of questions.

For this work, we are approaching peer review from more of a cognitive constructivist perspective. In particular, we will be using the theories of Bloom (1984) and Piaget (1950) to guide us. We are interested in how the student is affected by the acts of reviewing and being reviewed rather than by the social interaction occurring during the process.

Towards this goal, we are going to focus on a subset of Topping's topology. However, since we are looking at specific subsets and combinations of these dimensions, we are going to use our own terminology. This will allow us to refer more precisely to the portions we are interested in.

The dimensions we are exploring are interesting because they allow us to answer questions about the effects of *what* is being reviewed, *who* is being reviewed, and *when* they are being reviewed. What is being reviewed changes how the learners approach the review, what concepts they focus on, and what tools they use. The person being reviewed also has an effect. Whether it is another student or the instructor, who is being reviewed has a social impact on the process. The question of when lets us look at two aspects. First, the educational needs of freshmen and graduate students are very different. Peer review may be more or less effective in helping these students reach those goals and we should be aware of those differences. Second, reviewing can occur during or after the creation of a piece of work (term paper, program, etc.). Looking at and providing feedback on something when there is still time to make changes could have some obvious benefits. Similarly, being able to examine a completed and polished work could also be useful.

Because we using peer review in computer science classes, the *Product/Output* dimension is important to us. One of the issues here is use of design diagrams or code from programming projects in the reviews. Reviewing one of those items is not necessarily the same as reviewing a term paper, a musical score, a film, or anything else. To start with, students may have more or less familiarity with certain materials. Understanding how to read a paper is different from understanding how to read a diagram or code. Programs are not narratives and require different approaches to understand them. Review methods also change. For example, a student reviewing code can look at it statically or can compile and run it to view it dynamically. This restricts what we can infer from peer review experiences in other disciplines.  If we want to understand how reviewing programs helps students, we have to have them review programs.

Not only are reviewable objects in computer science different from other disciplines, they are different from each other.  When reviewing a programming project, two of the possible review opportunities are after students have done most of their designing, but before they have really started development, and at the end of development.  Both occasions have deliverables (diagrams or code) that lend themselves to review.  They, however, support different levels of inspection.  Design diagrams promote higher level, abstract reviewing while code can invite more detailed, low level examination.  As such, different learning objectives may be reached by using them.  We will refer to this section of Topping's *Product/Output* dimension as *materials*.

Other dimensions to explore are *Objectives* and *Directionality*.  Some of the possible reasons to perform a review include providing feedback, encouraging social interaction and learning, and giving students opportunities to learn how to evaluate.  In our work here, we are looking specifically at two kinds of reviews.  First, we are interested in peer reviews that have students dealing with other students and commenting on their work.  Second, we have training reviews designed to help the students learn to assess programming projects before they interact with their peers.

These two styles of reviews have effects on the *Directionality* used.  While *Directionality* also covers the selection of the reviewer and reviewee, we are focusing on whether a student is only a reviewer (training) or is both a reviewer and a reviewee (peer).  As mentioned before, giving and receiving reviews may have different benefits.  The level of social interaction may also affect the outcomes.  Being able to more clearly understand those benefits may lead to better designed assignments that take advantage of those differences.  The selection of reviewer and reviewee will be made randomly and will change over time.  This is done to provide the students with more exposure to other ideas.  Collectively, we will call these parts of the *Objectives* and *Directionality* dimensions the *type* of review.

The *Year* of the review is worth studying as well.  Topping uses the aspect to denote the relative class levels of the reviewer and reviewee.  Here, they both will be in the same class, so they will share the same *Year*.  What we are interested in the kinds of advantages or disadvantages there are in reviewing early or late in the curriculum or if it is even worthwhile.  The term *level* will be used for the *Year* dimension.

Deciding to review part way through or at the end of a programming project affects the *Focus* of the review.  If done at the design stage of a program, the reviews can be made into formative assignments as this would provide plenty of time to make changes to the project.  However, if the reviews are done at the end of the development or as a form of grading, the reviews may be more summative.  This is the *position* of the review.

There are also a number of dimensions that, for this work, remain relatively fixed.  Computer science is the *Curriculum Area*.  As the reviews will be done within a class, the *Ability* level will be roughly the same.  To simplify matters, the *Official Weight* and *Relation to Staff Assessment* will be kept

(as much as possible) the same between classes as will the *Requirement* and *Reward*. Since the peer reviews will be part of class they will be required and the students will be given some credit for their completion. The *Time* and *Place* for the reviews will generally be out of class but, may also be done as a lab exercise.

For *Privacy*, the reviews will be kept anonymous. (This is discussed more below.) Partially to protect anonymity and partially because of the digital nature of the materials being reviewed, the *Contact* will be online. The *Constellation Assessors* and *Constellation Assessed* will be fixed as a small group of individuals the size of which may vary with the size of the project.

### 2.1.3 Peer Review and the Student

One of the first things to do when changing how students are taught is to determine how it will affect students. Any change will be accompanied by benefits and detriments. Students may learn more from an extended time-on-task or they may become bored with it and consider it a waste of time. They could use the opportunity to develop their analysis and evaluation skills or they could use it to plagiarize code or to be overly critical of classmates. The possible effects need to be examined.

Davies and Berrow (1998) found students improved their self-evaluation skills through the process of evaluating the work of peers. There was also evidence that students felt that they were accountable for their efforts and that they could depend on the others in the class to participate in the process. There appeared to be little anxiety in making their work public. Another benefit found was that the students were able to determine what level of effort was acceptable because of the availability of other people's work.

Anewalt (2005) found reviews could improve the students' attitudes and self-efficacy. The process was designed to help the learners get comfortable with being evaluated in a professional setting. While there were some concerns raised, students mostly felt that the process was helpful in their learning and that it led to a higher quality project. Van Hattum-Janssen and Lourenco (2008) found related results in their work on peer and self-assessment.

Wolfe (2004) also describes benefits of the peer review process, such as students learning from peers and getting feedback of greater quantity than a teacher could reasonably provide. According to Wolfe, this allows students to develop their critical reviewing skills and to experience the other side of the grading process. This should make it easier for them to accept criticism. This also allows the professor to teach in a more constructivist manner (Ben-Ari 1988; Huba and Freed ; Kolari and Savander-Ranne 2000; McConnell 2001). The instructor is able to act as a guide rather than as the only source of knowledge. This can lead to the development of community and increase trust between peers. Wolfe found fewer problems than expected. Of the thousands of reviews, there were only a few cases where reviews were inappropriately negative or included personal attacks. Additionally, there was a strong positive correlation between participation in creating peer reviews and performance. Students reported that they were satisfied with the entire process.

In contrast to Wolfe, Davies (2000) found a greater number of problems with inappropriate reviews and with people upset with the feedback. He noted that students seemed to be willing to review other people's work, but they did not respond well to receiving reviews on their own work. While there were more favorable comments about the process than negative ones, it is still a potential problem to be addressed.  The design, the amount and type of training the students received before reviewing, and the motivation used may have caused the disparity in the amount of problems.  For example, Davies study had the students check each technical report for plagiarism.  Having the students actively looking for cheating could have put them into a mindset to be more critical and more negative.

Silva and Moreira (2003) suggest that peers can find problems that a busy instructor can miss.  This gives valuable feedback to the student and can improve the opportunity for growth.  However, if the peers are not experienced peer reviewers or are not very knowledgeable in the subject area, then the student may receive useless or invalid responses.  This can be a considerable worry to the students especially if the reviews count towards their grades.  There may be a need for the evaluation of the reviews, which may increase the effort required by the instructor.

Carroll and Rosson (1987) discuss a potential problem with the idea of a "production paradox."  The students' goal in taking the course is to learn the material presented and they enter the classroom with preconceived notions of how classes should run and how learning occurs.  So, there may be resistance when new teaching methods are applied.  Students may not be motivated to try new activities or to learn how to use any new tools because it takes time away from their current mode of learning that they are comfortable with and that they know will work for them.  While peer review and supporting tools may increase learning, it can appear to require too much effort or be too risky to take a chance on it.  Efforts need to be made to accustom the students to process and to familiarize them with the benefits so that they will buy into the activity.

### *2.1.3.1 Student Attitudes towards Peer Review*

Numerous studies have addressed the students' attitudes.  Some focused on the how well the students liked the program (Liu and Yuan 2003; Silva and Moriera 2003).  Some provided basic attitudinal information (Gatfield 1999; Purchase 2000; Gehringer 2003; Anewalt 2005)  while others provided more detailed views (Sivan 2000; Wen and Tsai 2006; Xiao and Lucking 2008) .  In general, the researchers found that students were accepting of the review process but some problems were found (Davies 2000).  This seems to be related to the manner in which the reviews were implemented.

### 2.1.4 Reliability of Review Scoring

Score or rating reliability is an issue that appears in much of the literature.  Many studies question whether students can produce ratings comparable to the instructors (Gehringer 2001; Liu, Lin et al. 2002; Hamer, Ma et al. 2005; Tseng and Tsai 2007; van Hattum-Janssen and Lourenco 2008).  There is always the concern, by the teacher and the students, that the reviews will be done poorly and that the misinformation will harm the students' grades.  This raises the question of whether, if reliability cannot be assured, peer feedback is an acceptable risk.  The results from the literature indicate that, while reliability is still an issue to consider, proper planning can circumvent many of the problems.  They (especially Falchikov and Goldfinch (2000)) also indicate that there is nothing inherent to computer science that would prevent peer review from being successful.  Some of these results are examined in more detail below.

In Topping's survey of the literature, he found, among other things, that studies that assessed the reliability of peer review found tremendous variations in their results (Topping 1998). He also notes that, since the purpose of teacher evaluation and peer evaluation might be different, then high reliability may not be required. That is, if the teacher acts as the grader and the students just provide feedback, then there is less need for reliable results as the students are not explicitly affecting another student's grade. That does assume that poor feedback from the reviewers does not lead the student into making incorrect decisions later.

Falchikov and Goldfinch (2000) found that reliability was related to the manner in which the review was designed. Their meta-analysis of 48 peer review studies found that well designed studies with well developed criteria for grading tended to be more reliable. As will be discussed later, the idea of using a good rubric may improve the reliability of the whole process. It is interesting to note that they did not find significant differences in reliability based on the level of the course or the discipline area. In general, the differences between single and multiple reviews were also not significant. However, a large number of reviewers seemed to hurt reliability. It was suggested that in larger groups, such as an entire classroom, there may be some "social loafing" and there may be less effort expended in doing the review.

Other work has tried to ensure reliability by computing a weighting for each reviewer. These weightings are then used to calculate the final score(s) that the reviewed student receives. One method for determining these weightings compares the ratings a student gave with the ratings from other students (Hamer, Ma et al. 2005). While this method appears to be effective, it does require that the students review a fairly high number (more than five and ideally ten) of assignments for the ratings to be useful. The Calibrated Peer Review system (Robinson 2001; Carlson and Berry 2005) uses a set of instructor provided sample assignments to train the students in the peer review process and to determine a reliability index. While this training may serve a very useful purpose, it is time consuming, for both the students and the instructor, and it needs to be repeated if the index is to be recalculated.

### 2.1.5 Quality of Peer Reviews

Aside from the reliability of the review, the quality is also an issue. Wolfe (2004) has found that novice graders can produce quite accurate assessments, but that a majority of the reviews could be classified as superficial. That is, they recognize if an assignment is good or bad but do not express constructive feedback. Partially, this could be overcome by additional training, as suggested by Topping (1998). Some other steps that could be taken to ensure quality reviews include the design of software that encourage teaching practices that can help elicit better responses from the students. This includes support for rubrics (discussed later) and the evaluation of reviews (Gehringer 2003; Sitthiworachart and Joy 2004).

Gehringer (2003) describes work with large classes where he has reviewers communicate with their reviewees through a shared web site. The system can be configured to allow or disallow reviewers to access the other reviewers' scores. While he says that this promotes more discussion about the review, he also mentions that the first reviewer's score might bias the others. Grading is based on a rubric with questions of weighted value. This improves the consistency of the grading since the rubric ensures that the students are evaluated on the same criteria.  The system also allows additional comments to be made to clarify the numeric scores.  In another effort to increase the quality of the reviews, Gehringer's system bases 25% of the grade on evaluations of peer reviews. He found students were motivated by this to review carefully.

Work by Chalk and Adeboye (2004), however, found a very different story.  Even when reviews count toward a quarter of the reviewer's grade, the researchers found no significant correlations between tutors' and students' evaluations in 4 of the 9 weeks.  This casts doubts on their use for actual assessment.  They suggest that peer evaluation be used as a learning exercise alone. Nonetheless, the researchers note evidence in the student reviews of higher level thinking skills of analysis and evaluation.

The quality of reviews can be quite a concern for students.  Following bad advice could lead to a lower score on an assignment.  Topping enumerates some methods to improve the quality of the process (Topping 1998).  These include familiarizing the students with the grading criteria, such as rubrics, checklists, or response grids, and providing exemplars to guide the students in the activity.  Related to that is the need to provide the students with enough training that they understand the procedure and that they know how to give and receive criticism.  Sufficient monitoring of the process by the instructor is also needed to help ensure a smooth process.  This may include the evaluation of reviews as one way to motivate students to do better critiques.  However, it does not seem that it is always the case.

### 2.1.6 Anonymity

Sullivan (1994) notes two paradigms for peer reviews, walkthroughs and inspections. With walkthroughs, reviewers are guided through the work by the reviewee, but with inspections, reviewers explore the material on their own without benefit of the reviewee's explanations.  For a type of walkthrough, Sullivan describes egoless teamwork, where peer review is reciprocal with small groups reviewing each others' work, to theoretically "level the playing field." But other researchers (Anderson and Shneiderman 1977; Cross 1987; Sitthiworachart and Joy 2004) note this ideal can be difficult to meet, and there is a need to model and monitor the outcomes of the reviews.  Many of the types of tools used to support peer review would typically take the form of Sullivan's inspections, since they can easy be done on-line and are more flexible in scheduling.  But, because the reviewee does not have an opportunity to present or defend his work to his peer, there may be greater vulnerability on the reviewee's part and a desire for anonymity.

Silva and Moreira's (2003) tool did not protect anonymity. In fact, students openly debated their work with their reviewers, which caused unnecessary friction among some students.  If students are not practiced in evaluation, they may be overly negative. Anonymity and instructor monitoring of the process are some methods of reducing anxiety and preventing hurt feelings from inappropriate criticism.  This is a similar approach to that used by Hundhausen, Agrawal et al (2009) only they did not seem to suffer from the negative consequences.  The design and implementation of the review may go a long way to preventing problems.

Double blind reviews, where that neither party know the others' identities, may promote fairer and better quality reviews (Zeller 2000). Zeller did not even allow instructors access to the reviews from the students. They noticed grade improvements and that the students using their system perceived a higher degree of fairness in the grading. However, the author does acknowledge that without instructor monitoring of the reviews there may be quality issues that arise.

The vast majority of the students Sitthiworachart and Joy (2004) interviewed thought anonymity (on both sides) was important. Wolfe (2004) used only one-way anonymity where the reviewer knows the author's identity, but not vise-versa. Gehringer (2003) and Chalk and Adeboye (2004) both use double blind anonymity. There are few explanations for any of these anonymity scenarios.

Depending on the goals for the peer review, anonymity has its uses as a way to remove some of the social pressure from the process. Students may be able to focus more on analyzing and evaluating if they are not worried about the social repercussions of their review. In addition, they may be more willing to accept the feedback they are given if they feel that it is not biased. Alternatively, not using anonymity may be desired for the social interaction and may be helpful in building communities.

As a final note, Davies and Berrow (1998) mention that anonymity is never certain. Students can include personal information in their documents or the program used to create the file could do it for them (i.e. Word's author information).

### 2.1.7 Assignment of Reviewers

Since assigning reviewers can become complicated and tedious, especially in a large class, there is a need to some way to handle this issue. Wolfe (2004) handled assignment by displaying how many reviews a student has received and directing students to review the work with the least number of reviews. This is probably not sufficient for most classes. Chalk and Adeboye (2004) used a random system where assignments were made on a rolling basis. Students could review others once they had submitted their work. This led to unbalanced numbers of reviews and instances where there was problems finding enough people to review. Topping (1998) mentions that peers may be matched in a variety of ways for a variety of reasons. They may be matched in groups or pairs by ability, friendships, or randomly. Different configurations support different goals. Randomly assigning reviewers to reviewees for each assignment may expose the students to a wider range of ideas. Matching students in pairs over a semester may not have the same breadth of ideas but allows for a longer term social interaction. The number of approaches to assigning reviewers suggests a flexible approach is needed to meet different needs.

### 2.1.8 Summary

There is a large amount of information in the literature about peer reviews and their effects. But, since the design space is so large, it is not enough. While some of peer review's benefits are known (Zeller 2000; Papalaskari 2003; Wolfe 2004; Hamer, Ma et al. 2005; Trytten 2005), it is not necessarily known how to achieve those benefits. The number of theories that could be used to (partially) describe peer review speaks to that. The literature discusses the process in terms from Bloom with an emphasis on cognitive processes (Bloom 1984) to Piaget's constructivist views (Piaget 1950) to Vygotsky and social interaction (Doolittle 1997). While those and many others are appropriate, they do not seem to cover the entirety of the subject (Topping 1998) and affect how the reviews are implemented and evaluated (Patterson 1996). These implementations will affect what is achieved during the process and how beneficial it is.

The discussions about the issues of student benefits, reliability, quality, and anonymity all contained different, sometimes conflicting, views of the problems and how to best address them. This is valuable but most of the arguments relate to the specific implementation and environment in which the studies were done. Reliability is a good example. There are a number of ways to improve or ensure that the students' evaluation of their peers is comparable to those of the instructor. However, this is only useful and needed if the reviews are used for grading other students. If reviewing instructor material or just providing descriptive feedback, reliability is not as great a concern.

What is not known is how to apply these ideas to other specific settings, like the computer science curriculum. There is not a good understanding of how the goals of engagement, high-level thinking, and conceptual learning fit with all of the other aspects of the review discussed here.

## 2.2 Rubrics

Sitthiworachart and Joy (2004) note that rubrics are important, especially in cases where students may not be comfortable or knowledgeable enough to create their own specific guidelines, which is probably true in most cases. They state that rubrics must be of a sufficiently large scale (5-point vs. 3-point) so that the students can mark effectively. Even so, they note that the process should be monitored.

Several of the systems reviewed (Mathews and Jacobs 1996; Gehringer 2001; Michael 2002; Trivedi, Kar et al. 2003; Sitthiworachart and Joy 2004; Hamer, Ma et al. 2005) provide for some kind of rubric-like functionality. Other parts of the literature made note of the need for them (Preston and Shackelford 1999; Guzdial 2001; Liu, Lin et al. 2002; Silva and Moriera 2003; McLuckie and Topping 2004; Trytten 2005; Hamilton, Brunell et al. 2006; Sadler and Good 2006; Wen and Tsai 2008). While there seems to be a general consensus that rubrics are important and that they improve the peer review activity, there is not as much agreement on how they should be implemented. Some just provide a way to select a numerical value (Joy and Luck 1998) while others provide for the addition of comments (Gehringer 2003). Still others give a list of predefined deductions and reusable comments that can be made (Mason and Woit 1999).

Work by Mayer (2004) suggests that guidance provided by rubrics may be very important to the learning process. In his paper, Mayer is specifically addressing the problem with using a pure discovery learning method but it has applications here. If students are not given a rubric or some similar set of marking guidelines, they must, of necessity, come up their own way to complete the assignment. There is no guarantee these "discovered" guidelines will be effective, useful, or correct which could severely compromise the benefits of performing a peer review. Mayer argues that providing some form of guidance to the process greatly improves the outcomes. The level of detail and completeness may vary greatly depending on the student's familiarity with the subject and the task, but rubrics or other instructions are important for their learning.

Sivan (2000) goes further than stating that a rubric is useful. The paper argues that the students need to familiar with and understand the rubric criteria so that they may review effectively. This may mean providing detailed descriptions of the criteria or, as Sivan suggests, engaging the students in their creation. By aiding in designing the criteria, they should become more involved and know what aspects (good and bad) to look as they review. Work by van Hattum-Janssen and Lourenco (2008) had similar success in giving students partial control over the definition of the grading criteria.

On the whole, the literature suggests that rubric support for reviewers is important and almost needed for novice reviews. Rubrics can supply the guidance students need to learn how to evaluate an assignment. It provides the needed scaffolding until the students are comfortable with the process and the domain to make correct judgments. The rubric and its criteria also define clear expectations for how students expect their work to be evaluated. This reinforces trust among all involved by allowing for more objective evaluation.

## 2.3 Education and Context

As has been noted in other areas, including communications, linguistics, education, and computer science (Suchman 1987; O'Grady, Dobrovolsky et al. 1989; Clark and Brennan 1991; Nardi 1996; Ryan 2001; Harrison, Tatar et al. 2007; Furberg and Arnseth 2009; Luehmann 2009; Teunissen, Stapel et al. 2009), context is extremely important. In this work, context refers to the environment in which the students are reviewing, including the class topics and review materials, the participants, and their expectations. As discussed with Topping's topology (1998), the *Product/Outcome* dimension affects the context in which peer review is performed. Different materials may require different approaches and have different effects. The same general argument could be made for several of the other dimensions but it is important for the *Curriculum Area* and *Year*. These two facets determine the concepts being learned in the class and influence the expectations of the students and instructors. Computer science students expect certain qualities from the discipline and they bring that with them (Seymour, Hart et al. 2004; Schulte and Magenheim 2005; Carter 2006). Understanding more about the elements surrounding the reviews is vital.

So, when engaging in the peer review process, this is something that needs considering. One needs to understand who is involved, know what their goals are, and determine what would constitute a successful review from their perspective. First, the roles of the participants in the review must be considered. Whether a person is reviewing or being reviewed makes a difference in what is important. For example, generating feedback may not be the top priority of the reviewer but it generally is very important to the one being reviewed. Creating a review process that does not require detailed feedback may accomplish a number of other goals, but it may not meet the reviewee's needs.

In the most basic terms, there are three parties associated with a peer review. They are:

- The reviewee

- The reviewer

- The instructor/administrator

These are similar in nature to the roles discussed by Wang, Yijun et al. (2008).

These are not necessarily exclusive roles (Topping 1998) but they have different goals and that is the next item to be considered. These three have a number of possible reasons to be involved with the review and these reasons range from practical aspects of the review to concerns about the social interactions that exists around it. Table 2 lists a few of these. Some of the goals work well together across participants. If the instructor tries to promote teamwork among a group, then that may encourage reviewers to help the reviewees by providing the detailed feedback they want. Then again, some goals may be independent of or in opposition to the others. A review activity designed to reduce the instructor's grading may not engage the reviewers enough to get them to analyze the work and provide useful or reliable comments. Knowing the objectives to support and possible interactions between them provides needed information for the design of a review.

**Table 2: Review Participants Goals**

| Participant | Possible Goals |
| --- | --- |
| **Reviewee** | To receive reliable, accurate reviews<br>To receive detailed feedback from multiple perspectives<br>To validate their own knowledge<br>To correct problems in their work |
| **Reviewer** | To display knowledge to others<br>To help others improve their work/learning<br>To understand how to evaluate work in this context<br>To learn from the work/ideas of others |
| **Instructor/ Administrator** | To reduce the amount of grading/oversight done<br>To promote learning in the reviewee and/or the reviewer<br>To promote social interaction and teamwork among a group<br>To give reviewers experience reviewing others<br>To give reviewees experience being reviewed<br>To generate resources (Gehringer 2003) |

Finally, knowing how to define success is important for a review. Simply put, the situation can radically change what is wanted out of a review. In industry, the goal may be to improve the product and enhance teamwork. Here, success may be fewer bugs and more communication. It may not be as important for the participants to learn new concepts as long as they can find potential problems. For an academic conference, a successful peer review may be one that takes the submitted papers and finds all of the high-quality ones that are relevant to its audience while weeding out the others. Although feedback is sent to the author, that is not necessarily the main purpose. In the classroom, reviews may be an effort to improve the learning of concepts (rather than simply making a project better) or of evaluation skills. Improved grades, better designs, more insightful reviews could all be signs that the goals have been accomplished.

### 2.3.1 Technology in Context

One idea that appears often in the literature is the use of technology to support the review process (Mathews and Jacobs 1996; Preston and Shackelford 1999; Davies 2000; Silva and Moriera 2003; Trivedi, Kar et al. 2003; Sitthiworachart and Joy 2004; Powell, Turner et al. 2006; Denning, Kelly et al. 2007; Hamer, Kell et al. 2007; Hou, Chang et al. 2007; Tseng and Tsai 2007; Wang, Yijun et al. 2008; Wen and Tsai 2008; Xiao and Lucking 2008). Since this idea is so common and since the additional of technology will change the setting in which the review is performed, it is worth discussing. First we talk about educational technology in general and then specifically about reviewing technology.

*2.3.1.1 Educational Technology*

The idea of technology in the classroom is not a new one. Computers, of one form or another, have been incorporated into the curriculum in many different ways. They have been used for specific functions such as data collection, visualizations, automated grading and note taking. In addition, they provide for more general features to increase communication among students and professors, to structure courses and their resources, and to allow for more learning opportunities (Jones and Johnson–Yale 2005). Or they can be used to allow for changes in pedagogy. Work by Minielli and Ferris (2005) suggest that the use of course management systems remove restrictions, in terms of location and time, which bind a classroom to a physical location and so change what and how ideas can be taught. Similarly, Moursund argued over thirty years ago that the introduction of calculators into the classroom provided for the opportunity to move away from teaching students how to perform calculations by hand and focus more on problem solving (Moursund 1976).

Part of the difficulty of introducing technology into the classroom is coping with unwanted side effects. Take, for example, an English class. The instructor's purpose may be to teach the students how to write a quality essay. There is no desire to teach the students how to use Word so that they can create an electronic document or, if the class is using some sort of course management system, to show them how to submit their files. Similarly, note taking is about capturing information, not about managing OneNote or other note taking programs. But these problems come up and distract from the true goal. Similar arguments can be made about most educational tools or, in fact, about most tools (Carroll 1997).

The field of Computer Supported Cooperative Learning (CSCL) (Koschmann 1996) explores this interaction between the student and the computer and its effects on learning. Part of this exploration is understanding what a student can accomplish with the technology and part of it is determining what lasting changes in cognitive processing have occurred. To clarify, a student may be taught concepts of object-oriented programming in a visual environment where drag and drop is used to connect and build objects. And, within this environment, the student might be able to build valid programs. However, when it is removed, the student may not transfer any of those concepts into another environment. So, it is important to know not only what is possible with technology but what learning will occur with its use.

Vrasidas (2004) raises a number of issues to consider about the use of technology in education. Specifically, he deals with e-learning, but there are wider applications. These concerns include:

- What are the changes in communication?

- What are the changes in the interaction and other social aspects?

- How does the value of the content change with the technology?

- What technologies are appropriate for what content?

As technology becomes more and more a part of the physical and on-line classroom, it will change how learning occurs and in what context.  It is vital to know what these changes are.

Tatar, Roschelle et al. (2003) and  Roschelle and Pea (2002) explored the use of handheld computers in classroom instruction.  While the handhelds were found to have valuable use in collaborative learning, they found a number of problems occurred as a result of the devices.  There were technical problems, such as how to synchronize data with a classroom full of handhelds, problems with usability issues when entering data, and problems with equipment damage.

In addition, they found instances of inappropriate use of the devices.  This brings up several concerns.  One problem deals with the way students and even professors use the technology.  It is reasonable to assume the some of the students will try to use the devices or software to play games, surf the web, or otherwise avoid doing the assignment.  It is hard to maintain the attention of a class when everyone has such a versatile device.  Vrasidas (2004) noted a similar problem with instructors.  When referring to learning management systems, that teachers did not necessarily use the systems as expected.  The basic versatility of the computer could become a hindrance here.

Cheating is another trouble.  Being able to share data and collaborate within groups also allows for inappropriate sharing.  This is of special concern for peer review in computer science classes.  Students are taught to make modular and reusable code.  Since the peer review activity provides students with other people's assignments, this makes it especially easy and tempting for students to take pieces of code and to directly use them in their own work.

In their work on Presenter, Anderson, Anderson et al. (2004), found related issues.  In their analysis of the use of digital ink in lectures, they noticed that the users did not take advantage of many of the available features (Anderson, Hoyer et al. 2004).  Presenter gives the user basic pen features such as multiple colors, different ways to erase the screen, and highlighter functionality.  While these seem to be obvious and very useful options, some of them were not used much in practice.  Instead of using the highlighter, instructors would use marks to draw attention to the material.  Changes in pen color generally occurred when it was necessary, such as to ensure proper contrast.  This lack of use may be because they are not essential to the activity.  Switching pen colors is a small disruption, but it interrupts the task.

Work by Turner, Pérez-Quiñones et al. (2006a) shows some of the effects that technology has on educational processes.  When students were asked to review three UML diagrams, one using paper and pen, one using a tablet PC, and one using a desktop, the students produced significantly more comments on paper than they did on either the tablet or the desktop.  While this does raise some questions about how technology should be used in that type of situation, it is important to note how widely the responses varied.  Although paper was the overall favorite, every platform had some people who preferred it, sometimes heavily, over the others.  The reactions were very personalized and it should be understood that people will interact with technology differently.  Similar ideas were explored in the context of note taking and with similar results. (Turner, Kim et al. 2006).

Tatar, Roschelle et al. (2003) and Carroll (1997) offer some advice for how to handle these types of issues.  Tatar emphasizes enhancing rather than replacing.  That is, try to create technology that addresses very specific issues and not a system that does everything for every possible situation.  In the classroom, that means supporting existing practices and not replacing those that already work well.

A minimalist approach to the design of the systems is also needed (Carroll 1997).   Students should be able to get access to a tool and immediately use it for a real task.  For example, if a student must spend a lot of time to learn how to submit files or to use commenting tools, then the software is inadequate.  The tool should focus on the user's needs rather than on what might be occasionally useful.

In general, tools, and specifically educational tools, should be designed knowing that they will have unintended effects.  Designing with this in mind will help produce more usable systems.

### *2.3.1.2 Peer Review and Technology*

There are two basic types of technology useful to the peer review process.  One type aids in the administration of the review.  This includes providing functionality for distributing and collecting materials, assigning reviewers to reviewees, and supporting the use of rubrics.  Most of the systems in the literature reviewed here are of this type.  The other kind of technology allows for the annotation of the materials being reviewed.  Examples of this include Microsoft's Word and OneNote programs that support flexible document markup.

The need for administrative tools becomes apparent when we consider that we want students to be focused on reviewing and learning rather than on the management papers and files.  For instance, students may be using development environments, such as Eclipse (2007), to do their coding.  So it would be useful to allow them to submit their projects directly to be peer reviewed instead of having to find the source files, zip them together, and then submit them through email or to a course management system.

For the instructors, there is also a practical need to provide support for setting up and performing peer reviews (Mathews and Jacobs 1996; Papalaskari 2003).  Many classes that commonly use peer review, such as composition courses, are, generally, relatively small with 20 to 30 students.  In the first two years of the computer science curriculum (CS1 and CS2), programming classes may be, and frequently are, much larger and this makes the administration significantly more difficult.  Assigning reviewers to reviewees and ensuring the proper transfer of artifacts between them becomes increasingly more difficult as the number of students rises.  This becomes even worse if the instructor wants to try to monitor the quality of the reviews or to provide some anonymity to the students.

So, there is a need to remove some of this burden from the professors so that it becomes reasonable to assign peer review activities.  Simply put, if it is easier to do, it is easier to do more frequently and the more exposure the students have to this task, the more they will be comfortable with it and the more they will learn from it.  Software could be and is used to help gather in submissions, distribute them to reviewers and then to do the same thing in reverse with the reviewed assignments (Davies 2000; Liu and Yuan 2003; Silva and Moriera 2003; Trivedi, Kar et al. 2003; Wolfe 2004; Carlson and Berry 2005; Hamer, Kell et al. 2007).  Other tools can support the instructor in creating and maintaining rubrics (Joy and Luck 1998; Gehringer 2001; Sitthiworachart and Joy 2004; Ross-Fisher 2005)and sets of grading instructions.  These rubrics provide students with enough guidance to allow them to review without having to develop their own criteria of what a good project is.

It is important, however, to consider the purposes of these tools.  If they are developed to aid instructors by moving the burden of handling the details onto the students, then they may end up doing more harm than good.  Likewise, if the administration overshadows the actual purpose of the exercise then it becomes a waste of time.  These tools should be evaluated and developed with the students and the review goals in mind.

If that idea is kept in mind, there is no reason that computer-aided or online peer reviews cannot be successful.  Cross (1987) compares three methods of peer review: face-to-face, computer mediated, and merging peer reviews together. He found that students were able to independently create useful comments and that each of the three methods produced comparable reviews, demonstrating that alternatives to face-to-face methods can be equally useful.

One tool designed to aid in the administration of review exercises is a peer review module (Powell, Turner et al. 2006) created for Moodle (Moodle.org 2005), an open source course management system.  This module allows the instructor to assign peer review activities to students.   At the moment, assignments and resources available through Moodle can be reviewed.  That is, the system can handle the review of material generated by the class and any materials uploaded by the professor, such as sample projects or work from previous classes.  The instructor can create, reuse, and share rubrics to be used in the activity and/or can require students the students to upload an annotated document containing their review.  The rubric supports multiple choice, multiple answer, and true/false questions and allows for an exemplar to be added for each value in a criterion (A good design has the following features…).  It also supports comment boxes after each question.

All reviews, both rubric results and documents, are available to the teacher who can assign a grade to them.  In addition, the professor has control over the assignment of reviews.  They can be assigned randomly, manually, or the assignments from another peer review activity can be used.  Among other things, the instructor can control whether the reviews are anonymous or not.  If they are anonymous, the names of both the parties are hidden from the students but not from the teacher.

The students are provided with a screen listing the reviews they are to complete and of reviews their work received.  The system links in the appropriate files and provides access to an instance of the rubric (if used).

This system was created with many of the design guidelines, identified in the literature, in mind. The goal of the system was to provide support and flexibility in the creation and administration of the process. This includes helping the instructor create and publish rubrics for the students (and instructor) to use. We used this system to support many of the activities in our current work.

As with the administrative tools, it is easy to see that there is a need for tools that easily allow annotation of the artifacts being reviewed. Ideally, one should be able to open coding projects or design diagrams and just start adding comments. As with the administration of files, the focus should be on the reviewing and not on converting files to different formats or otherwise using awkward methods to add reviews. A system similar to that described by Allowatt and Edwards (2005) would be useful in this regard.

Another example is minimUML, a program developed to aid in the markup of UML design documents (Turner, Pérez-Quiñones et al. 2006b). minimUML is educational tool that takes a minimalist (Carroll 1997) approach to UML diagramming. In this, it is of similar design and purpose as Alphonce and Ventura's QuickUML (Alphonce and Ventura 2002; Alphonce and Ventura 2003). The tool, which is intended to support the typical amount of UML used in CS1 and CS2 classes (Lewis and Chase 2005), allows students to quickly create and publish (to an image file, to a printer, or as basic, skeleton code) basic UML diagrams but it also supports peer review with two forms of annotations: typed "post-it" notes and freeform ink drawings. Possible future work on the program includes the possibility of adding more review features such as the ability to have multiple reviewers on a single document (and be able to control their display) and the integration with Web-CAT, an automated grading system (Edwards 2003).

### 2.3.2 Summary

After considering Topping's topology (1998), it becomes clear that the setting, the materials being reviewed, the participants and their expectations all interact with the peer review process. Understanding the review involves understanding all these other influences as well. This includes the students and their opinions about what is and what is not computer science. This includes the instructor and the goals for the class. This includes the technology the students and teacher use in the learning activities. Being able to fit the peer review into that context will improve its usefulness and make it more successful.

# 3 Exploration of Peer Review in Computer Science

## 3.1 Summary

Overall, the purpose of this study was to explore the topic of peer review in computer science and to identify ways it affects the reviewers and develop guidelines for creating review exercises.  We explored a number of avenues and collected a large amount of data.  Unfortunately, the data collected was spotty, to say the least, and was not linked together well enough to support a reasonable, detailed analysis to meet our goals.

One issue we discovered was that our design was too broad to give definitive results.  We did, however, have enough data to allow us to narrow down the focus for a second study.  From the analysis of one class, we were able to identify two interesting areas for further research.  The *type* of review appears have a significant effect on the length and focus of the review.  We also found evidence that students reviewed some of the concepts differently than they did others.  Both of these findings should be explored in the future work.

## 3.2 Goal

The goal of this study was to explore how peer review affects students as it is used in a computer science classroom.  From the literature, we have some information on how it is used in other disciplines (Falchikov and Goldfinch 2000; Liu and Hansen 2002; Dossin 2003; Carlson and Berry 2005) and we have some theories (Piaget 1950; Bloom 1984; Doolittle 1997; McConnell 2001) that can be used to help explain the process.  However, there is not a clear understanding of the *materials* being reviewed, the *type* of review, how the *level* of the class, or the *position* of the review affect the benefits to reviewers in the various programming classes.

As described above, we decided to explore the differences in the review process by varying a number of dimensions.  These were based on the topology provided by Topping (1998) and tailored to our specific circumstances.  Our dimensions (with Topping's corresponding ones added in parentheses) included:

- *Materials*  being reviewed (Product)
    - Programming code
    - Design diagrams
- *Type* of review (Objective, Directionality)
    - Training reviews with material provided by the teacher
    - Peer reviews with material generated by classmates
- *Level* of the class (Year)
    - From CS1 and CS2 to the Graduate level

- *Position* of the review (Focus)
    - Early in project development (Formative)
    - Late in project development (Summative)

We felt that varying these aspects of the review assignments would provide us with a rich set of data to explore and would allow us to identify trends and develop some guidelines.

## 3.3 Methods

### 3.3.1 Participants

Participants were students enrolled in eight computer science programming courses. Activities in the study were presented as part of their normal class work. Depending on the class, the number of participants ranged from roughly 10 to 60 although there was a great amount of variability in the number of those that completed all parts of the study. An assortment of students both computer science majors and non-majors and from freshmen to graduate participated. Again, this depended heavily on the class.

### 3.3.2 Experimental Design

For each class, the students were presented with the following assignments:

- Pre-questionnaire
- Training exercise(s) on teacher provided material
- Peer review exercises(s)
- Post-questionnaire

The exceptions to this format were class 2-F06 which was only given the training exercises (see Table 3 for the class ids and more information).

Eight computer science programming courses implemented the peer review activities for this study. For each class, the reviews were designed to support the educational goals of the class. For example, in class 1-SII07, a CS1 class for non-majors, exercises focused on coding while 3-F07, a CS3 class, focused on design. For each review, the students were asked to review a small number of their peers (2-4 based on the size of the project) using a rubric to guide them (Mayer 2004; Sitthiworachart and Joy 2004; Hamilton, Brunell et al. 2006) (See Appendix A: Class 3-F07 Review Rubric). It was intended that every class would complete at least one training review and at least one peer review during the semester. For a number of reasons, discussed below, this did not occur for all of the classes. For one class, the position of the reviews within the project was varied as a way to explore its effect on the process. This occurred in class 1-F07.

The reviews, the questionnaires, and anonymous grade information were collected for analysis.

*3.3.2.1 Questionnaires*

Students received a questionnaire before the peer review activities and then a similar questionnaire at the end of the semester.  They are made available online through a local survey hosting system.  The questionnaires cover topics that included:

- Need for cognition (short form) (Cacioppo, Petty et al. 1984)

- Perceptions of working with peers

- Value of peers' input

- Self-efficacy (Perceived skill level compared to the class)

- Fears about peer review

- When they typically start a programming project and how they manager their time on task

- Demographic information (non-identifying)

The questionnaires were intended gauge the students' attitudes towards peer review, their self-efficacy, their motivation for when they start on their projects, and their typical work patterns as they complete programming projects.  Parts of the survey are similar to survey one by Wen and Tsai (2006).  The need for cognition (Cacioppo and Petty 1982) questions help highlight some of the internal motivation for working on their project.  The second questionnaire did not contain the need for cognition questions and asked about their work patterns on the peer reviewed projects.  No identifying information was collected in the surveys. (See Appendix B:  Questionnaires.)

*3.3.2.2 Rubrics*

Rubrics were provided for all review activities.  These varied slightly from class to class in order to meet the needs of the class.  The rubrics included some of the following dimensions with the first five being core to most.

- Functionality (Does the assignment meet the requirements of the specifications?)

- Abstraction (How well is abstraction used?)

- Decomposition (How well is the problem divided up and put together?)

- Encapsulation (How well is information protected?)

- Style (Does the work meet style and documentation guidelines?)

- Similarity/Novelty (How does this work compare to your own?)

- Testing (How well is the code tested by the included student tests?)

The categories we used were partially based off on rubrics found in the literature (Joy and Luck 1998; Trivedi, Kar et al. 2003; Sitthiworachart and Joy 2004; Trytten 2005; Denning, Kelly et al. 2007; Sanders and Thomas 2007).  We took the common concepts indentified and grouped them into more general categories.  The result was a set rubric criteria covering high-level concepts which could be broadly applied to CS projects.  (See Appendix A: Class 3-F07 Review Rubric.)

The categories of Similarity/Novelty and Testing were added for specific purposes.  Similarity/Novelty was used explicitly to encourage students to compare their own work to the material they were reviewing.  We felt that self-reflection was a valuable portion of the review process and wanted to provide an opportunity for it.  This same idea was also covered in the post-questionnaire.  The Testing category was used in classes being taught with a test-first approach (Beck 2001).   In seemed appropriate to have students evaluate the effectiveness of the tests created for the project.

For each dimension, multiple levels of achievement were provided.  Again, the wording and number of levels changed somewhat between classes but the level typically ran from "Excellent" to "Poor" and, for each level, a description of what it meant to be at that level was provided.  These descriptions were designed to guide students in their reviews by reemphasizing the definitions of the dimensions and placing value on particular actions.

In the instructions given to the students, they were asked to compare the project to the exemplars contained in the rubric, to select a rating for that category and then to provide a detailed rationale.  We tried to stress the need for describing why a rating was given and to encourage them to be as thorough as possible.

None of the rubrics were designed to be checklists.  That is, no specifics were given in what classes, methods, variables, algorithms, or data structures should be present.  The rubrics were created to encourage students to think on a conceptual level and not to "grade" the work.

### 3.3.3 Data Collection

The data collected for this study did not occur as smoothly as we would have wished.  Several factors limited the amount of data we gathered.  These included difficulties with scheduling the exercises in an existing class, with students completing only some of the reviews or questionnaires, with low numbers of students enrolling in a particular course, and with the abnormal completion of one of the semesters in which the study was being run.  As a result, we were able to collect a large amount of data but it is not as complete as we would have liked.

#### 3.3.3.1 Questionnaires

The number of responses to the questionnaires was low.  Three classes had no post-questionnaire responses at all.  Of those classes that did have responses for the second questionnaire, the number was too small to lend any confidence to a statistical analysis (See Table 3).  As a result, we looked at the data from the surveys as a way to guide future work.  We found that, while not directly answering our questions, the information about their attitudes towards peer review and their internal motivation to think may be needed clarify some of the results we found while analyzing the rubrics data.

Since we did not have any way of associating the questionnaire answers with any of the other data collected, we did not pursue an analysis towards that end. This will be discussed more below with the data analysis of the rubric data.

### 3.3.3.2 Rubrics

The collection of data from the reviews was more successful. From the eight classes, we were able to record 996 completed rubrics from 299 reviewers. While the amount of data was large, it is also incomplete. Of those classes which were intended to have both training and a peer reviews, three of them were not able to complete the second review assignment and, so, have nothing to be compared to. Two of the other classes have only a moderate number of participants (15-20) which is not as high as we would have liked for our statistical analysis. One class produced no viable information at all (See Table 4 for details).

**Table 3: Collected Data**

| Class ID | 1-SII07 | 1-S07 | 1-F07 | 2-F06 | 2-S07 | 2-F07 | 3-F07 | G-S07 |
|---|---|---|---|---|---|---|---|---|
| Course | CS1 (non-major) | CS1 | CS1 | CS2 | CS2 | CS2 | CS3 | Grad Level |
| Semester | Summer II 07 | Spring 07 | Fall 07 | Fall 06 | Spring 07 | Fall 07 | Fall 07 | Spring 07 |
| # Pre-questionnaire | 14 | 17 | 10 | - | 57 | 63 | 35 | 10 |
| # Post-questionnaire | 12 | - | 9 | - | - | 0 | 11 | 4 |
| Training Type | Small code exercises | Code | Code | Code | Code | ? | Design | - |
| Training | In class exercises | Review of DUML | Review of DUML | Review of PS Interpreter | Review of PS Interpreter | ? | Review of Movie Database design | - |
| Review Type | Code | - | Code | - | - | ? | Design | Technote |
| Review | Review of project 5 and 6 | - | Peer review of web objects | - | - | ? | Review of Cruise Control System | Review of Technote 1,2,3 |
| Grades | Yes | Partial | Yes | Yes | Partial | No | Yes | Partial |

**Table 4: Assignment Details**

| Class ID | Type | Assignment Reviewed | # of Reviewers | # of Objects/Peers | Total Reviews | # of Rubric Categories |
|---|---|---|---|---|---|---|
| **1-SII07** | Peer | Project 5 | 14 | 3 | 42 | 4 |
|  | Peer | Project 6 | 11 | 3 | 33 | 4 |
| **1-S07** | Training | DUML Translator | 66 | 3 | 198 | 6 |
| **1-F07** | Training | DUML Translator | 44/43/43 | 3 | 130 | 6 |
|  | Peer | Web Objects | 37 | 2 | 74 | 5 |
| **2-F06** | Training | PS Interpreter | 61 | 2 | 122 | 6 |
| **2-S07** | Training | PS Interpreter | 50/48 | 2 | 98 | 6 |
| **2-F07** |  |  | 0 | 0 | 0 | 0 |
| **3-F07** | Training | Cruise Control System | 45 | 2 | 94 | 6 |
|  | Peer | Movie Database | 34 | 3 | 101 | 6 |
| **G-S07** | Peer | Technote 1 | 19 | 2 | 28 | 10 |
|  | Peer | Technote 2 | 19 | 2 | 38 | 10 |
|  | Peer | Technote 3 | 19 | 2 | 38 | 10 |
| **Total** | **Peer:** 7 **Training:** 5 **Multiple reviews:** 4 |  | **299** |  | **996** |  |

### 3.3.4 Data Analysis

For this study's data analysis, we focused on the information generated in our CS3 class (3-F07). We chose this class because it had largest, most complete set of data. We were able to record data on both a training and a peer review with a reasonable number of participants in both. This class also had the best response to the questionnaires. With this set of data, we are able to explore the *type* of review but not the other three dimensions as we had intended. Our analysis was designed to highlight interesting features in the data to explore in more depth in a later study.

As mentioned, the 3-F07 class completed both a training review and a peer review. The training review required the students to review designs based on the specifications for a cruise control system task that was given as homework in a previous offering of the class. Each student reviewed the same two designs. These designs were based on student work that was modified and anonymized to meet our needs. Specifically, we wanted to provide the reviewers with a "good" and a "bad" design. Both of these artifacts had some strong and some weak points but they did not have anything that was obviously and horribly deficient or that was clearly created by someone with many more years of experience. Besides the designs, students were provided with the original assignment information before they started their review. In total, 94 completed rubrics were collected and 68 of these were analyzed.

In preparation for the peer review, the students were asked, as homework, to create a UML diagram for a movie database system. This was a design problem similar in nature to other design problems at this level. The specification provided opportunities for students to use inheritance, aggregation, and association and to make several design choices. A reasonable solution to the assignment could be composed of 6-10 classes.

Each student was assigned to review three movie database designs created by their peers. Reviewers were randomly assigned to reviewees. The review was anonymous and conducted online through an extension to our course management system (Powell, Turner et al. 2006). From this, we collected 101 completed rubrics, all of which were analyzed.

From both reviews, we were able to evaluate 34 students' work. We excluded those students that did not complete one or more rubrics in both assignments. We did, however, include those students that provided no comments in their reviews (i.e. only selected a rating.). See Table 5 for details.

**Table 5: 3-F07 Data**

| Type of Review | Reviews per student | Design Task | Reviewers | Rubrics Collected | Rubrics Analyzed |
|---|---|---|---|---|---|
| **Training** | 2 | Cruise Control System | 45 | 94 | 68 |
| **Peer** | 3 | Movie Database UML | 34 | 101 | 101 |
| **Total** | 5 | | | 195 | 169 |

### 3.3.4.1 Data Coding

The individual rubric categories were coded for each rubric. In the 3-F07 class, the rubric contained six categories. These were:

- Abstraction
- Decomposition
- Encapsulation
- Functionality
- Style
- Novelty

After a brief analysis of the rubric data, we decided not to continue analyzing all of the categories. The data in the Style, Novelty, and Functionality categories did not appear to hold much information of value. For this particular set of assignments (two designs) given to the students, there was not a large amount of style to consider. Responses tended to be along the lines of "It looks fine to me" or "This should be improved" without anything else of real depth. The comments for Novelty were in a similar vein. While having them answer the question may have required some self-reflection and provoked some learning, the comments were generally pretty shallow. Functionality also tended to have banal answers.

So, our discussion will focus on the concepts Abstraction, Decomposition, and Encapsulation. Each category was coded using a set of five measures (see below). As mentioned above, students completed multiple rubrics per assignment. There were two reviews for the training exercise and three for the peer. Responses by the same student in the same exercise were coded separately. These values were aggregated in the statistics, but it allowed us to see more variation in the way that the students reviewed. We also calculated summary data for the entire rubric.

Focusing our analysis on the rubric enables us to see a number of things. As a whole, the rubric can be an indicator of how the reviewer interacts with the peer review process and may show where it is working or failing. By analyzing the rubric we can get an indication of how seriously the task is being taken. Whether the students think of it as busy work or as something useful should be recognizable in their work. Also, by measuring changes, we can see if they are learning something.

Specifically, by coding the quality of the completed rubrics, we should be able to see how well the students:

- Understand the concepts involved in the review
- Are able to think at higher levels
- Engage with the topic

Positive changes in the quality between review offerings would be an indication of learning.  Negative changes could show lack of engagement.

The rubric can also show where process is working, failing or encouraging certain behaviors.  If we find consistently bad reviews across the entire class, then we must reconsider the implementation.  Students may not understand the process, may not have the necessary skills to evaluate well, or may find it a waste of time.  Whatever the outcome, this provides more information about how to create review assignments that work even if it is just what not to do.

### *3.3.4.2 Measures*

Based on what we can draw out from the quality of the rubrics, we designed our measures to cover three things:

- Understanding of concepts
- Level of thinking
- Engagement

Getting students to *understand concepts* is a major goal of computer science programs.  It is not enough to have students that can write code.  They must be able to decompose problems, create coherent representations and abstractions and do so in a clean, safe, encapsulated manner.   The application of these concepts in a review gives us a view of the students' understanding.  Improvement in the understanding of these concepts shows learning.

Encouraging students to develop a high *level of thinking* is one of the goals of higher education.  We want to have students working on the higher levels of Bloom's taxonomy (Bloom 1984).  Problem solving requires those higher-level skills and that is an integral part of computer science (Gagne, Briggs et al. 1992; Trytten 2005).  Measuring how well the students perform on a task requiring those higher levels of thought gives us insight into the students' ability to function on those levels.

*Engagement* is a basic part of active learning and other constructivist theories (Ben-Ari 1988; Ertmer and Newby 1993; Doolittle 1997; Kolari and Savander-Ranne 2000).  We want students to take more interest, more control of their learning.  If we have activities that provide a challenge, offer noticeable benefits, or otherwise draw their attention, we can increase their time-on-task and the amount of effort they put into it.  High levels of engagement are good indications that learning will follow.  It is not a guarantee, but it is much more likely than if they are not engaged.

All together, these three aspects show an effective constructivist environment.  It indicates that students are active, that they are being required to use and build their knowledge (rather than regurgitating it), and that the activity promotes learning.  Positive measurements of *engagement*, *level of thinking*, and *understanding of concepts* is a sign of success.  As we have defined success, that means that the process is beneficial to the reviewers.  On the other hand, if these measurements are not as good as we would want, then it shows us where to improve our instructional designs to better serve the students.

The measures we used in our analysis were:

- Four steps of reviewing
    - Measuring *level of thinking*, *engagement*
- Level of review
    - Measuring *level of thinking*, *engagement*
- Match of comment to criteria category
    - Measuring *understanding of concepts*
- Detail of response
    - Measuring *understanding of concepts*, *engagement*
- Length of reviews
    - Measuring *engagement*

### 3.3.4.2.1 Four Steps of Reviewing

It is theorized that to be successful reviewers, students must go through four steps during the process (van den Berg, Admiraal et al. 2006). There are:

- Analyze
- Evaluate
- Explain
- Revise

In the Analyze step, the reviewers are looking through and trying to understand the work. Evaluation is making a decision on whether it meets the standards or not. The third step, Explain, involves providing details as to why it does or does not meet those standards. Finally, suggestions for improvements or other approaches happen in the Revise step.

In a good review, these four steps should be evident. It is a sign that the student is ready for this type of activity and that the review assignment is appropriate. The presence of these shows that the reviewer was *engaged* enough to be able to explain problems found and possible solutions. It also displays that the student could take part of the *higher-level thinking* activities of analysis and evaluation.

The presence and relative quality of these steps indicates the quality of the review. Positive changes in that quality may show learning and higher level-thinking. Since this is a learning process, we are not expecting perfect reviews. It has been shown that novices do not use the steps as systematically as experts do (van den Berg, Admiraal et al. 2006). However, if the exercise is working well, then the quality should improve over time as the students have more practice with the skill.

If steps are consistently missing or degrade in quality, then the review may not be functionally well. It may not be engaging enough. Or, the students may not have a strong enough background in computer science to be able to really know what to look for. Another possible explanation is that the students are missing some required skill or knowledge. They may just not be familiar enough with the review process to do it well. More time and training may be required. It may also be an indication that the students are not developmentally ready to think on this level (Perry 1970; Bloom 1984).

In the coding of this measurement, we looked for the presence and the quality of the step. The responses for each of the criteria were labeled with one or more of the 4 steps (i.e. A, E, X, R, AE, AX, …, AEXR). The comment was marked with N/A if none of the steps could be reasonably applied to the comment. To look at the quality of the step, we looked at how consistently it appeared in the student's remarks.

### 3.3.4.2.2 Level of Review

Since one of our purposes is to prompt students to think at higher levels, we are interested in how abstract or concrete their comments are. In English composition, it has been found that novice reviewers tend to make lower level changes than experts (Sommers 1980; Flower, Hayes et al. 1986). In general, novice writers did not change meaning but made more superficial changes in spelling and grammar or structure. They focused on conventions and rules (Sommers 1980; Nold 1982). We assume that these general finding carry over into computer science and that neophyte reviewers will direct their comments to low level details and will not suggest changes to the overall design.

This level of review gives us a view of the students' *level of thinking*. If the comments are abstract then they are likely to be working on Bloom's Synthesis, Analysis, or Evaluation levels. If they are more concrete, they may be working on the lower Knowledge or Application skills. More abstract remarks may also show more *engagement*. To make strong, abstract statements about a project, especially when one is describing the decomposition, one needs to understand the bigger picture and understand how the parts are intended to work together. That requires an investment of time and effort in the review.

To code the level of review, we assigned a number to the comment based on the detail it contained. The score ranged from 1 to 5. If an observation focused on design, algorithms, etc. then it was assigned a 1. If it concerned syntax, the naming of variables or methods, etc. it was assigned a 5. Where comments had a mixture of both abstract and concrete elements, they were rated by the relative amount of each type of comment. A value of N/A was assigned to those statements that did not relate to the material (e.g. "Nice job!"). An average was calculated for the student for each assignment. This is in addition to the average for the rubric.  These values allow us to compare changes in student performance between reviews.

### 3.3.4.2.3 Match of Comment to Criteria Category

In addition to reviewing well and abstractly, the students need to be able to address the topic at hand.  The match of the comment to the criterion is one way to measure the *understanding of the concepts*.  If they are focusing on other topics, it could be that they do not know enough about the concept to evaluate it well.  Or they may misunderstand the concept partially or entirely.  Changes in the match of the comment may show conceptual learning.

We should also consider the social context of the review in this case.  The match of the comment may be influenced by the students' desire to provide as much information as possible to their peers regardless of its fit to the category.  On the other hand, students may limit their comments and narrow their focus in order to not look foolish to their classmates.   This is a confounding factor that we will discuss in the results of the study.

To code the match of the comment, we assigned a number from 1 to 5 to each remark based on how well it addressed the category and only that category.  A 1 was assigned if the comment did not address any of the concepts in that category.  A 5 was used for comments focused entirely on the criterion.  Other values were assigned based on the relative ratio of related comments to unrelated comments.  If no comment was made, a value of N/A was used.   Besides the average calculated for the rubric, averages, for Abstraction, Decomposition, and Encapsulation were computed for the student for both assignments.  These values allow us to compare changes in student performance between reviews.

### 3.3.4.2.4 Detail of Response

A complement to both the level of review and the match of the comment is the level of detail in the response.  We want them to explain their thinking because it gives them a chance to reflect on their ideas.  It also is an opportunity for instructors to identify misunderstandings the students have (Turner, Quintana-Castillo et al. 2008).  In either case, we prompted the students in the instructions to provide more than a "this is good" kind of answer.  They were specifically asked to provide a rationale for their choices.

With this setting we can use this measure to gauge both *understanding of concepts* and *engagement*.  It stands to reason that students who are more comfortable with their knowledge may be more verbose.   Those who do not have a good handle on the topic may write less because they have less to talk about.  However, like the match of the comment, this too may be influenced by social pressure. Students may feel the need to provide more details to fellow students to help them out.  They may also comment more on the materials to get a better grade from the instructor.  Alternatively, they may be less detailed if they do not want to offend their peers by pointing out mistakes or if they do not want to look ignorant on the topic.  Again, this is a factor we will discuss with the results.

As a way to show *engagement*, it is reasonable to assume that students who are more detailed in their comments spent more time thinking about the material and were more involved. Very brief and vague comments can be seen as a way for a student to complete the assignment as fast as possible.

The coding for the detail of response is similar to those for the level of review and the match of the comment. We assigned a number from 1 to 5 based on the level of detail displayed. A 1 was given for a very superficial comment or one with no explanation at all. Very detailed, very specific remarks were given a 5. An N/A was used where there was no comment made. Once again, we calculated an average for the student for each assignment. These values allow us to compare changes in student performance between reviews.

### 3.3.4.2.5 Length of Reviews

The length of the comment was used as one way to estimate *engagement*. While not a very clear measure, since there can be a lot of variation between students, it does provide an idea of the effort the student put in. More writing would show that more work was done. For this measure, we simply counted the number of words in each comment. Averages were created for the rubric and for each student per assignment. Changes in length between assignments could show a change in *engagement*.

# 4 Results

With the coded data from the 169 rubrics from class 3-F07, we performed our analysis. Calculations were done with JMP and SAS.

## 4.1 Analysis of the Four Steps of Reviewing

To analyze of the four steps of reviewing data, we used the a GEE (generalized estimating equations) model (Liang and Zeger 1986). We had tried a number of other categorical analysis models that produced similar results. The GEE, however, seems to most appropriate method for this particular set of data.

While looking for differences between the training reviews and the peer reviews, we found no significant differences in the student use of the reviewing steps (Analyze, Evaluate, Explain, and Revise) (See Appendix C: GEE Model SAS Code and Results). It should be noted that we were looking for the presence of these steps in discrete occurrences. We did not look at the steps in combination. That is, a comment coded as both Analyze and Explain would be viewed as a comment that had the Analyze step and one that had the Explain step but not as one with an Analyze/Explain step.

We did, however, find a difference in the way students reviewed the concepts. There were significant differences between the Abstraction, Decomposition, and Encapsulation categories in the use of the review steps ($p<0.05$). This holds true for each of the four steps.

From this we can see that students reviewed these concepts differently. This may be because the students understand these concepts at different levels. It could be that the average student learns these three concepts at different rates. For instance, Abstraction may be harder to conceptualize than Encapsulation. From this, we may be able to show that, in this context, that peer review is not as effective for some concepts as others. Students may not be ready to evaluate some of the concepts because they do not have enough prerequisite knowledge.

Another explanation is that it may require a different level of effort to review the concepts (e.g. Decomposition may be harder to see in code than in a diagram.) This could change how we plan and use review activities. For example, if it is easier to see Decomposition in a UML diagram, then a design review is needed to promote that concept.

If reviewers are having trouble with some of the concepts, then they may not be learning much as they try to evaluate their use in a project. We need to be able to understand this difference to adequately plan the creation and implementation of review exercises. Unfortunately, appealing to the literature is not of much use. There does not seem to be much pervious work that directly addresses how students' understanding of these concepts develops.

## 4.2 Analysis of the Level of Review

For this portion, we started by taking the difference between the students' average score on both reviews. We did this for all three categories and for the rubric average. We then used a paired t-test and the signed rank test on the data. The data appears to be normally distributed.

We found no significant differences in any of the areas ($p>0.1$). While we wanted to see a progression to more abstract, higher-level thinking, this was not the case. Students reviewed on roughly the same level on both reviews. It is likely that two review activities are just not enough of an intervention to prompt a noticeable change. This may come with more experience.

We did find a correlation with the measurements for the detail of response. Using Spearman's ρ (1904), we found that they were correlated at approximately 0.64 ($p<0.0001$). We say approximately because we found this correlation in all three rubric categories with slightly different values (See Table 6). This indicates that the more concrete comments also tended to be more detailed. This is rather intuitive as it is very easy to provide specific information about the low-level aspects of the project, such as the naming conventions or stylistic concerns.

Table 6: Correlation of Level of Review and Detail of Review

| Category | Spearman's ρ | p |
|---|---|---|
| Abstraction | 0.64 | < 0.0001 |
| Decomposition | 0.67 | < 0.0001 |
| Encapsulation | 0.64 | < 0.0001 |

## 4.3 Analysis of the Match of Comment to Criteria Category

As we did with the level of review, we took the difference between the students' average score on both reviews assignments. This was done for all three categories and for the rubric average. We then used a paired t-test and the signed rank test on the data. The data appears to be normally distributed.

We found significant differences in all four cases (See Table 7). As explained above, this shows a change in how well the students tailored their comments to the concept in question. In this case, the students performed better on the peer review (the second exercise). They stayed on topic better and this may show a change in their understanding of the concepts.

Table 7: Change in match of comment to criteria category

| Category | p<t |
|---|---|
| Abstraction | <0.001 |
| Decomposition | <0.05 |
| Encapsulation | <0.05 |
| Rubric | <0.001 |

So, this could be the indication of learning we were looking for. It demonstrates that the reviewers can at least separate out one concept from another and it may also show that they are evaluating better. Unfortunately, there is a wrinkle in this analysis. The value of this data depends greatly on the students' motivation. It could be that they understand the concepts and the process better and can, therefore, comment more precisely. Or, it could be that they are trying to avoid work or social awkwardness by limiting their remarks and saying less. We do not have the information in this study to make that determination. We need more information about the students' attitudes towards the peer review process and their motivation to participate in it. That would give us more data with which to disambiguate these results.

It is also unclear whether these differences are due to the *type* of review or the *order* in which it was given. This requires more detailed research. If it is the *type*, then we have a reason to use one implementation over the other. If the review of their peers elicits more on topic discussions, then that is an advantage over the training reviews. It may also indicate that fewer exercises are needed to produce some of the results we want. On the other hand, if it is an order effect, then the repeating the process is enough to improve results. That will allow us to choose the *type* of review based on other criteria. It also suggests that more repetitions may enhance results further.

## 4.4 Analysis of the Detail of Response

Once again, we took the differences between the students' average scores for all three categories and the entire rubric on both reviews assignments. We then used a paired t-test and the signed rank test on the data. The data appears to be normally distributed.

We did not find any significant differences in the data (p>0.1). Since the students did not seem to change how detailed their comments were, there was no evidence of a change in their *understanding of the concepts* or in their *engagement*. From looking at the data, it appears the some students may be engaged in the task but a number of others are not. Perhaps there needs to be more external motivation for this type of assignment.

As with the level of review, we found some interesting correlations. Using Spearman's ρ (1904), we compared looked at the interactions between the concept categories. We found that Abstraction was moderately, but significantly, correlated with both Encapsulation and Decomposition (p<0.05) (See Table 8). There was no correlation between Decomposition and Encapsulation (p>0.1). This is an interesting finding as it suggests that understanding Abstraction may help with understanding the other two concepts. That may suggest an order in which those concepts are taught. It may also relate to the difference we found in how the concepts were reviewed. This requires more research to clarify the finding.

**Table 8: Correlations in the Level of Detail**

| Categories | | Spearman's ρ | p |
|---|---|---|---|
| Abstraction | Encapsulation | 0.44 | <0.05 |
| Abstraction | Decomposition | 0.35 | <0.05 |
| Decomposition | Encapsulation | - | >0.1 |

## 4.5 Analysis of the Length of Reviews

When examining the length of the comments, we took the differences between the students' average scores for all three categories and the entire rubric on both reviews assignments. We then used a paired t-test and the signed rank test on the data. The data appears to be normally distributed.

We found significantly different lengths between the two assignments in all the cases except for Decomposition (See Table 9). The reviews completed in the training exercise (the first assignment) were longer by, on average, eight words. This may show a (negative) change in engagement over time. It is also noteworthy that the difference in Decomposition was not significant while the other concepts were. This is further evidence that students review and understand the concepts differently.

Like with the match of the comment, we have multiple ways of interpreting this data. It could possibly be a sign that the students are more *engaged* and *understand the concepts* better and, therefore, are not putting in some of the filler they used in the first assignment. On the other hand, it could be that they are less *engaged* and are just not working as hard. Once again, we have the need for attitudinal and motivational information. That should help clarify the situation.

There is an order effect to consider here as well. If the *type* is the influence, students may be trying to be more verbose for their instructor or social pressure may make for briefer answers. If order, it may be that the students are tiring of the process or think that they can get away less work than they put in the first time around. Depending on the reason here, there is an effect on how to design reviews. If it is the *type*, then using training reviews may make the students freer with their comments. If order, additional work may be needed to keep the students interested in the reviews. More research is required.

**Table 9: Change in Length of Comment**

| Category | p<t |
|---|---|
| **Abstraction** | <0.05 |
| **Decomposition** | <0.1* |
| **Encapsulation** | <0.05 |
| **Rubric** | <0.01 |

**\* not significant**

# 5 Thoughts

We have learned a few things from this data. First, we found differences between the two review exercises. We found that the peer review exercise had a better match of comment to criteria category. We also saw that the training reviews had longer comments on average. These findings tell something about the *engagement* and the students' *understanding of concepts* but both of them could be a positive or a negative indication of change. More information about the attitudes of the reviewers is needed to determine which it is. This information is part of what we need to make decisions about how or if we use review assignments of this nature in this context.

Second, we saw differences in how some of the concepts were reviewed. Using the four steps of reviewing, we noticed variations between Abstraction, Decomposition, and Encapsulation. This is also hinted at in the length of the comments. We may have identified concepts that are learned at different rates or a situation where there is a dependency on other concepts. This has both implications on how those concepts are taught and on how they are used in reviews. Reviewing may be better for some of the concepts than others at this level. If the reviewers are having more trouble with some of the concepts, then that may hurt their engagement and their general attitudes towards the review process.

## 5.1 Remaining Questions and Follow-up

While this data did not give us definitive answers, it helped us narrow down the questions. We now have a stronger foundation to start exploring and, from there, we can plan how to extend the research.

### 5.1.1 Type versus Order

- Is it *type* or order (or both) that is causing the effects the training and peer review?

As discussed, students commented more directly on the criteria in question but wrote shorter comments for the peer review. We need to explore these variations and see if training and peer reviews elicit different responses. We have several possible explanations for the differences. Students may feel social pressure and be reacting to that. They might be focusing more on the feedback because they know it is helping a classmate or they might be writing more in the hope that something they put down is what the teacher wants. It also could be that, by the second review, the students are more experienced reviewers and understand the material better so they can write more on-point, concise reviews. An alternative is that they are getting lazy and writing less.

From this, instructors can plan exercises more effectively. It might be useful to have only training or only peer reviews. It there is an order effect, then having several reviews during a semester may be a good path. For one, it may change the reviewers' experience. They may enjoy (or dislike) working with their peers. That will affect their level of *engagement* and their attitudes. While we have been focusing on the reviewer in our work here, in the larger picture, teachers must also balance the reviewee and administrative needs as well. For instance, training reviews are easier to set up and control but they do not help the reviewees. Knowing if the *type* has an effect on the reviewer, gives more information about the choice.

### 5.1.2 Concepts
- Why are there differences in how concepts are reviewed?
    - Are there differences in conceptual difficultly?
    - Do the reviews improve student learning of these concepts?

There may be learning implications that are related to the differences we found between the criteria of Abstraction, Decomposition, and Encapsulation. All three concepts require advanced mental development (Piaget 1950; Perry 1970). Abstraction involves thinking in vaguely defined terms. Decomposition requires thinking about an entire project and exploring how to can be split, merged, and understood. Encapsulation calls for the evaluation of the benefits and detriments of protecting portions of the implementation. If students are learning and understanding these concepts at different rates, then we should review our teaching methods and practices. If one concept that is noticeably harder or easier than the others, it may give us ideas about where to place instructional emphasis.

There is little in the literature that covers this type of distinction or order effect in object-oriented (OO) concepts. In linguistics, research has found that beginners tend to learn morphemes (arbitrary signs) in certain orders (O'Grady, Dobrovolsky et al. 1989). There may be a similar issue with OO concepts. If we understand that ordering, we can arrange instruction accordingly.

We must also have a way to gauge student understanding of these concepts. The concepts of Abstraction, Decomposition, and Encapsulation are core to the object-oriented computer science curriculum but it can be difficult to tell if they are learning and understanding these ideas. We have already seen that students can perform well in a class while carrying significant misunderstandings about the topics, so other viable methods are needed (Turner, Quintana-Castillo et al. 2008).

Just as important, we want to know if completing reviews is a good way to enhance their understanding of these concepts. If not, then we might want to try a different approach. Our current approach used a high-level rubric. We asked about concepts and, while we gave guidance, we made the questions open in terms of how they were applied. Perhaps we are not asking the right questions or are on the wrong level. We may want to modify the focus of the review or change the how we divide the concepts used in the rubric criteria.

### 5.1.3 Interest and Attitudes
- Is reviewing an engaging and interesting task in computer science?

One aspect of our goals in this work is to create an environment in which students can actively learn. As part of that, we should know if what we are doing is drawing students in the activity. This should be an activity students are willing to participate in and not one that we are forcing on them. We need an assessment of their interest in and feeling about reviewing. It is important that we look at these attitudes over time for improvements or for problems to be addressed.

### 5.1.4 Benefits

- Are there significant learning benefits to reviewing in the early computer science curriculum as compared to other, common homework/lab exercises?

While we have identified a number of potential benefits from reviewing, we have not shown that it is better than or as good as what we currently do.  We require some sort of baseline to compare our efforts to.   We need a control group in our experiments in order to judge effectiveness.

## 5.2 Summary

The nature of this study has been exploratory in nature and, from it, we have learned more about how peer review functions in the computer science classroom.  Our original goals included the exploration of this process along the dimensions of the *materials* being reviewed, the *type* of review, the *level* of the class, and the *position* of the review.  Unfortunately, because of the amount of incomplete data we collected, we could only focus on one class and the effects of the *type* of review.

There were a couple of noticeable outcomes from our data analysis.  First, there were differences between the training review and the peer review in terms of how well student comments matched the criteria category of the rubric and in the length of the comments.  We also found differences in the way that the students reviewed the concepts of Abstraction, Decomposition, and Encapsulation indicating that they have different levels of understanding for these concepts or are learning them at different rates.  Our data analysis also highlighted the need for attitudinal information to help explain some of the results.

With those findings in mind, we developed questions for further research.

- Is it *type* or order (or both) that is causing the effects the training and peer review?
- Why are there differences in how concepts are reviewed?
    - Are there differences in conceptual difficultly?
    - Do the reviews improve student learning of these concepts?
- Is reviewing an engaging and interesting task in computer science?
- Are there significant learning benefits to reviewing in the early computer science curriculum as compared to other, common homework/lab exercises?

These will allow us to define the goals and the direction of further exploration into the *type* of review and into conceptual learning.  Additional future work involves the exploration of the *materials*, the *level*, and the *position* dimensions that we were not able complete with this study.

# Appendix A: Class 3-F07 Review Rubric

## Functionality
Does the code support the required functionality?

| | |
|---|---|
| Excellent | It is evident that all the functionality required by the specification is available.  It is easy to see which methods and which classes will perform which actions. |
| Good | All the functionality seems to be there but there is some question about how everything will work together.  It should work well if not perfectly. |
| Satisfactory | Most of the functionality is there but it may not work as well as planned.  It is not immediately evident that it conforms to the specification and may fail for certain tasks. |
| Unsatisfactory | Some things may work, but there seems to be a lot of functionality missing. Starting over from scratch may be to right thing to do. |
| Not acceptable | Was this based on the right specification? |

## Abstraction
How well is abstraction used?

| | |
|---|---|
| Excellent | Each class conforms to a single clear abstraction.  Every method is appropriate to the class and does not detract from the abstraction. Everything makes sense. |
| Good | Each class relates to an abstraction but there is an occasional exception.  A few methods may break the abstraction or a couple of classes may not from a coherent entity.  It could be improved but not too much. |
| Satisfactory | Abstractions are used but it is not rare to find places where they do not hold. Classes may attempt to be two things at once and may have methods that do not relate at all.  There is some confusion about what the classes are attempting to represent. |
| Unsatisfactory | Abstractions are extremely weak or nonexistent.  Methods are not relevant or break the abstraction.  Very little is coherent. |
| Not acceptable | Everything was put together randomly. |

## Decomposition
How well does the code use inheritance, association, aggregation?

| | |
|---|---|
| Excellent | Everything is broken down into appropriate objects.  Inheritance is used to take advantage of commonalities between classes.  Functionality is built and achieved through the use of aggregation and association.  Everything has a clear purpose and place. |
| Good | Aggregation, association, and inheritance are used well throughout.  There are a few points where inheritance could have been used better or where classes could have been divided or combined for better effect, but most decisions are reasonable and defendable. |

| Satisfactory | Some decomposition has been used but there are several missed opportunities.  Classes with common features do not use inheritance appropriately and some classes could, very reasonably, be restructured.  More thought could be used here. |
| Unsatisfactory | Classes tend to be monolithic when they are easily dividable.  Classes duplicate functionality in other classes when inheritance is clearly needed.  This was done haphazardly. |
| Not acceptable | It's all one big class. |

## Encapsulation

How well is the code encapsulated?

| Excellent | Everything that should be private is private and everything that should be public is public.  Access to variables is given only as needed.  Values that should not be changed cannot be changed.  All needed constructors are there and not more. |
| Good | Almost everything is encapsulated properly.  There are a couple of cases where variables are accessible where they should not be, values can be changed when they should not, or there are missing or inappropriate constructors.  It needs a little work, but not much. |
| Satisfactory | A fair attempt has been made to encapsulate everything, but there are several cases where it fails.  Most values can be accessed and modified whether or not it is appropriate.  It would be good to reevaluate all of the constructors as well.  This could be tightened up a lot. |
| Unsatisfactory | Not much thought has been put into encapsulation.  Variables may be public, everything is accessible and modifiable.  The constructors are not very useful.  Most classes need to be heavily reworked. |
| Not acceptable | Everything is public. |

## Style

How well does the code meet style and documentation guidelines?

| Excellent | The names of variables and methods are all consistent and self-explanatory, so that their role is clear without requiring reference to comments or documentation. |
| Good | Most names are well-chosen, but a small percentage of the user-defined names do not meet the requirements for excellence and there is some room for improvement. |
| Satisfactory | A solid attempt to choose meaningful names has been made, but there is significant room for improvement. Many names could be refined. Some methods may be given misleading or uninformative names and some data members may give little or no indication of the role the data plays. |
| Unsatisfactory | Although some names are acceptable, many names violate expectations, are poorly chosen, or hamper readability. |

| Not acceptable | No attempt has been made to use well-chosen names or to follow acceptable naming conventions. |

## Novelty

Were there any new techniques, tricks, or concepts that you noticed in the code that seem like they could be useful in future projects?

| **Many** | Yes, there were many ideas that will be incorporated into future projects. |
| **Some** | Yes, there were several ideas that could be useful or a few ideas that will definitely be helpful. |
| **Few**  | Yes, there were a couple of ideas that look like they could be useful. |
| **None** | No, there was nothing that looked promising. |

This rubric is partially based off a program grading rubric created by Dwight Barnette and used in the CS 2605: Designing Data Structures and OO Software I class at Virginia Tech in the Fall 2007 semester.

# Appendix B: Questionnaires

## Pre-questionnaire

I would prefer complex to simple problems.

    extremely uncharacteristic
    somewhat uncharacteristic
    uncertain
    somewhat characteristic
    extremely characteristic

I like to have the responsibility of handling a situation that requires a lot of thinking.

    extremely uncharacteristic
    somewhat uncharacteristic
    uncertain
    somewhat characteristic
    extremely characteristic

Thinking is not my idea of fun. *

    extremely uncharacteristic
    somewhat uncharacteristic
    uncertain
    somewhat characteristic
    extremely characteristic

I would rather do something that requires little thought than something that is sure to challenge my thinking abilities. *

    extremely uncharacteristic
    somewhat uncharacteristic
    uncertain
    somewhat characteristic
    extremely characteristic

I try to anticipate and avoid situations where there is likely a chance I will have to think in depth about something. *

    extremely uncharacteristic

somewhat uncharacteristic
uncertain
somewhat characteristic
extremely characteristic

I find satisfaction in deliberating hard and for long hours.

extremely uncharacteristic
somewhat uncharacteristic
uncertain
somewhat characteristic
extremely characteristic

I only think as hard as I have to. *

extremely uncharacteristic
somewhat uncharacteristic
uncertain
somewhat characteristic
extremely characteristic

I prefer to think about small, daily projects to long-term ones. *

extremely uncharacteristic
somewhat uncharacteristic
uncertain
somewhat characteristic
extremely characteristic

I like tasks that require little thought once I've learned them. *

extremely uncharacteristic
somewhat uncharacteristic
uncertain
somewhat characteristic
extremely characteristic

The idea of relying on thought to make my way to the top appeals to me.

extremely uncharacteristic
somewhat uncharacteristic
uncertain

somewhat characteristic
extremely characteristic

I really enjoy a task that involves coming up with new solutions to problems.

extremely uncharacteristic
somewhat uncharacteristic
uncertain
somewhat characteristic
extremely characteristic

Learning new ways to think doesn't excite me very much. *

extremely uncharacteristic
somewhat uncharacteristic
uncertain
somewhat characteristic
extremely characteristic

I prefer my life to be filled with puzzles that I must solve.

extremely uncharacteristic
somewhat uncharacteristic
uncertain
somewhat characteristic
extremely characteristic

The notion of thinking abstractly is appealing to me.

extremely uncharacteristic
somewhat uncharacteristic
uncertain
somewhat characteristic
extremely characteristic

I would prefer a task that is intellectual, difficult, and important to one that is somewhat important but does not require much thought.

extremely uncharacteristic
somewhat uncharacteristic
uncertain
somewhat characteristic

extremely characteristic

I feel relief rather than satisfaction after completing a task that required a lot of mental effort. *

extremely uncharacteristic
somewhat uncharacteristic
uncertain
somewhat characteristic
extremely characteristic

It's enough for me that something gets the job done; I don't care how or why it works. *

extremely uncharacteristic
somewhat uncharacteristic
uncertain
somewhat characteristic
extremely characteristic

I usually end up deliberating about issues even when they do not affect me personally.

extremely uncharacteristic
somewhat uncharacteristic
uncertain
somewhat characteristic
extremely characteristic

I enjoy working with my peers in a computer science class.
strongly disagree	disagree	neutral	agree	strongly agree

I enjoy working with my peers but not in a computer science class.
strongly disagree	disagree	neutral	agree	strongly agree

I do not like working with peers.
strongly disagree	disagree	neutral	agree	strongly agree

I feel that my peers do not have the knowledge to give me valuable feedback on my programming.
strongly disagree	disagree	neutral	agree	strongly agree

I feel my peers can help me improve my programming.
strongly disagree	disagree	neutral	agree	strongly agree

I feel my peers can help me improve my design for a program.
strongly disagree    disagree    neutral    agree    strongly agree

I feel that my peers can give me valuable feedback on the coding of my program.
strongly disagree    disagree    neutral    agree    strongly agree

I feel that my peers can give me valuable feedback on the design of my program.
strongly disagree    disagree    neutral    agree    strongly agree

I like choosing the student with whom I'll be working.
strongly disagree    disagree    neutral    agree    strongly agree

I feel the professor should assign pairs of peers according to abilities.
strongly disagree    disagree    neutral    agree    strongly agree

I feel my programming skills are better than most of the other students in this computer science class.
strongly disagree    disagree    neutral    agree    strongly agree

I feel my knowledge of the syntax for the language used in this class is better than most of the other students in this computer science class.
strongly disagree    disagree    neutral    agree    strongly agree

I feel that my design skills are better than other students in this computer science class.
strongly disagree    disagree    neutral    agree    strongly agree

I am a better than average programmer.
strongly disagree    disagree    neutral    agree    strongly agree

Peer reviews are a waste of time.
strongly disagree    disagree    neutral    agree    strongly agree

Peer reviews will help me improve my grade on the program.
strongly disagree    disagree    neutral    agree    strongly agree

I'm afraid my peer will give me "incorrect" feedback (for example, ask me to change my design structure incorrectly).
strongly disagree    disagree    neutral    agree    strongly agree

I do not feel comfortable reviewing other students' programs.
strongly disagree    disagree    neutral    agree    strongly agree

I enjoy looking at other students' programs.
strongly disagree	disagree	neutral	agree	strongly agree

The professor's feedback is more useful than my peer's feedback.
strongly disagree	disagree	neutral	agree	strongly agree

The TA's feedback is more useful than my peer's feedback.
strongly disagree	disagree	neutral	agree	strongly agree

The value of a peer review depends on the peer with whom you have been paired.
strongly disagree	disagree	neutral	agree	strongly agree

The following questions ask about how you work on projects in your Computer Science courses. Please respond to them with that context in mind.

Normally, I start working on a project
Immediately after it is assigned.
Well before it is due, but not immediately.
About a week or so before it is due.
A few days before it is due.
Right before it is due.

Normally, I have the majority of my project finished
Well before it is due
Shortly before it is due
On the day it is due
A couple of hours before it is due
After it is due

In general, I TRY to work on my project in (pick all that apply)
Many short sessions (1-3 hours long).
Several longer sessions (4-6 hours long).
A few long sessions (6+ hours long).
One marathon session.

In general, I ACTUALLY work on my project in (pick all that apply)
Many short sessions (1-3 hours long).
Several longer sessions (4-6 hours long).
A few long sessions (6+ hours long).
One marathon session.

In general, I find that
I have plenty of time to finish my project.
I have adequate time to finish my project.
I am somewhat rushed to finish my program on time.
I am rushed to finish my program on time.
I am very rushed to finish my program on time.
I do not finish my program on time.

In general, I start working on a project because
I enjoy the challenge.
I want to get it over with.
I have free time at that moment.
I have other work due about the same time.
It is due soon.

Please select your gender.
Male
Female

Please indicate your connection to Computer Science
Declared Computer Science Major
Declared Computer Science Minor
Considering Majoring or Minoring in CS
Taking this course for fun

What grade do you expect to get in this course (realistically)?
A
B
C
D
F

# Post-questionnaire

I enjoy working with my peers in a computer science class.
strongly disagree	disagree	neutral	agree	strongly agree

I enjoy working with my peers but not in a computer science class.
strongly disagree	disagree	neutral	agree	strongly agree

I do not like working with peers.
strongly disagree	disagree	neutral	agree	strongly agree

I feel that my peers do not have the knowledge to give me valuable feedback on my programming.
strongly disagree	disagree	neutral	agree	strongly agree

I feel my peers can help me improve my programming.
strongly disagree	disagree	neutral	agree	strongly agree

I feel my peers can help me improve my design for a program.
strongly disagree	disagree	neutral	agree	strongly agree

I feel that my peers can give me valuable feedback on the coding of my program.
strongly disagree	disagree	neutral	agree	strongly agree

I feel that my peers can give me valuable feedback on the design of my program.
strongly disagree	disagree	neutral	agree	strongly agree

I like choosing the student with whom I'll be working.
strongly disagree	disagree	neutral	agree	strongly agree

I feel the professor should assign pairs of peers according to abilities.
strongly disagree	disagree	neutral	agree	strongly agree

I feel my programming skills are better than most of the other students in this computer science class.
strongly disagree	disagree	neutral	agree	strongly agree

I feel my knowledge of the syntax for the language used in this class is better than most of the other students in this computer science class.
strongly disagree	disagree	neutral	agree	strongly agree

I feel that my design skills are better than other students in this computer science class.
strongly disagree   disagree   neutral   agree   strongly agree

I am a better than average programmer.
strongly disagree   disagree   neutral   agree   strongly agree

Peer reviews are a waste of time.
strongly disagree   disagree   neutral   agree   strongly agree

Peer reviews will help me improve my grade on the program.
strongly disagree   disagree   neutral   agree   strongly agree

I'm afraid my peer will give me "incorrect" feedback (for example, ask me to change my design structure incorrectly).
strongly disagree   disagree   neutral   agree   strongly agree

I do not feel comfortable reviewing other students' programs.
strongly disagree   disagree   neutral   agree   strongly agree

I enjoy looking at other students' programs.
strongly disagree   disagree   neutral   agree   strongly agree

The professor's feedback is more useful than my peer's feedback.
strongly disagree   disagree   neutral   agree   strongly agree

The TA's feedback is more useful than my peer's feedback.
strongly disagree   disagree   neutral   agree   strongly agree

The value of a peer review depends on the peer with whom you have been paired.
strongly disagree   disagree   neutral   agree   strongly agree

The following questions ask about how you worked on projects in this course that were peer reviewed. Please respond to them with that context in mind.

In general, I started working on a project
Immediately after it was assigned.
Well before it was due, but not immediately.
About a week or so before it was due.
A few days before it was due.
Right before it was due.

In general, I had the majority of my project finished
Well before it was due
Shortly before it was due
On the day it was due
A couple of hours before it was due
After it was due.

In general, I TRIED to work on my project in (pick all that apply)
Many short sessions (1-3 hours long).
Several longer sessions (4-6 hours long).
A few long sessions (6+ hours long).
One marathon session.

In general, I ACTUALLY worked on my project in (pick all that apply)
Many short sessions (1-3 hours long).
Several longer sessions (4-6 hours long).
A few long sessions (6+ hours long).
One marathon session.

In general, I found that
I had plenty of time to finish my project.
I had adequate time to finish my project.
I was somewhat rushed to finish my program on time.
I was rushed to finish my program on time.
I was very rushed to finish my program on time.
I did not finish my program on time

In general, I started working on a project because
I enjoyed the challenge.
I wanted to get it over with.
I had free time at that moment.
I had other work due about the same time.
It was due soon.

Please select your gender.
Male
Female

Please indicate your connection to Computer Science
Declared Computer Science Major
Declared Computer Science Minor

Considering Majoring or Minoring in CS
Taking this course for fun

What grade do you expect to get in this course (realistically)?
A
B
C
D
F

# Appendix C: GEE Model SAS Code and Results

Analysis provided by Mari Rossman and Dr. Ying Liu as a part of the Laboratory for Interdisciplinary Statistical Analysis (LISA) program.

```
proc genmod data=new;
class type ID class step_A;
model step_A =type class/dist=binomial  link=logit type3;
repeated subject=ID/type=cs;
run;
```

```
  The GENMOD Procedure

                    Analysis Of GEE Parameter Estimates
                    Empirical Standard Error Estimates

                              Standard    95% Confidence
       Parameter      Estimate   Error       Limits            Z Pr > |Z|

       Intercept       -0.1465  0.2068  -0.5519   0.2589  -0.71   0.4787
       Type    Peer     0.1894  0.2485  -0.2976   0.6765   0.76   0.4459
       Type    Training 0.0000  0.0000   0.0000   0.0000    .       .
       Class   Abst     0.5825  0.1856   0.2187   0.9463   3.14   0.0017
       Class   Deco     0.2139  0.2298  -0.2365   0.6643   0.93   0.3518
       Class   Enca     0.0000  0.0000   0.0000   0.0000    .       .

                    Score Statistics For Type 3 GEE Analysis

                                     Chi-
                   Source     DF    Square    Pr > ChiSq

                   Type        1     0.57       0.4503
                   Class       2     8.05       0.0179
```

<span style="color:red">The class significantly affect on Step_A. Review Type does not significantly affect on Step_A.</span>

```
proc genmod data=new;
class type ID class step_E;
model step_E =type class/dist=binomial  link=logit type3;
repeated subject=ID/type=cs;
run;
```

  The GENMOD Procedure

Analysis Of GEE Parameter Estimates
                       Empirical Standard Error Estimates

                                Standard    95% Confidence
       Parameter         Estimate   Error       Limits         Z Pr > |Z|

       Intercept           0.1807  0.2218  -0.2541   0.6154   0.81   0.4154
       Type     Peer      -0.1549  0.2104  -0.5673   0.2574  -0.74   0.4615
       Type     Training   0.0000  0.0000   0.0000   0.0000    .       .
       Class    Abst      -0.0712  0.1849  -0.4335   0.2912  -0.38   0.7003
       Class    Deco       0.4893  0.2358   0.0272   0.9514   2.08   0.0380
       Class    Enca       0.0000  0.0000   0.0000   0.0000    .       .

                   Score Statistics For Type 3 GEE Analysis

                                    Chi-
                 Source      DF    Square   Pr > ChiSq

                 Type         1     0.54      0.4644
                 Class        2     6.25      0.0439

```
proc genmod data=new;
class type ID class step_X;
model step_X =type class/dist=binomial  link=logit type3;
repeated subject=ID/type=cs;
run;
```

The GENMOD Procedure

                    Analysis Of GEE Parameter Estimates
                       Empirical Standard Error Estimates

                                Standard    95% Confidence
       Parameter         Estimate   Error       Limits         Z Pr > |Z|

       Intercept          -0.4035  0.2374  -0.8688   0.0618  -1.70   0.0892
       Type     Peer       0.2668  0.2593  -0.2415   0.7750   1.03   0.3036
       Type     Training   0.0000  0.0000   0.0000   0.0000    .       .
       Class    Abst      -0.2461  0.2085  -0.6547   0.1626  -1.18   0.2379
       Class    Deco       0.4295  0.2157   0.0067   0.8523   1.99   0.0465
       Class    Enca       0.0000  0.0000   0.0000   0.0000    .       .

                   Score Statistics For Type 3 GEE Analysis

                                    Chi-
                 Source      DF    Square   Pr > ChiSq

                 Type         1     1.04      0.3072
                 Class        2     6.62      0.0365

```
proc genmod data=new;
class type ID class step_R;
```

```
model step_R =type class/dist=binomial  link=logit type3;
repeated subject=ID/type=cs;
run;
```

```
              The GENMOD Procedure

              Analysis Of GEE Parameter Estimates
                Empirical Standard Error Estimates

                          Standard    95% Confidence
   Parameter      Estimate   Error       Limits         Z Pr > |Z|

   Intercept       2.1218   0.3170   1.5005   2.7431   6.69   <.0001
   Type    Peer    0.0101   0.3065  -0.5905   0.6108   0.03   0.9736
   Type    Training 0.0000  0.0000   0.0000   0.0000    .      .
   Class   Abst   -0.7846   0.2995  -1.3717  -0.1976  -2.62   0.0088
   Class   Deco   -1.1437   0.2919  -1.7157  -0.5716  -3.92   <.0001
   Class   Enca    0.0000   0.0000   0.0000   0.0000    .      .

              Score Statistics For Type 3 GEE Analysis

                                Chi-
            Source       DF    Square    Pr > ChiSq

            Type          1     0.00       0.9736
            Class         2    11.26       0.0036
```

The class significantly affects on Step_A, Step_E Step_X and Step_R. Review Type does not significantly affect on Step_A, Step_E Step_X and Step_R.